\documentclass[preprint, 5p, towcolumns]{elsarticle}
\usepackage{amssymb}
\usepackage{amsmath}
\usepackage{lineno}
\usepackage{xcolor}
\usepackage{float}
\usepackage{pifont} 
\usepackage{siunitx} 
\usepackage{url}
\usepackage{soul}
\graphicspath{{figures/}}

\begin{document}

	\begin{frontmatter}
		
		\title{Breakdown Performance of Guard Ring Designs for Pixel Detectors in $150~\mathrm{nm}$ CMOS Technology}
		
		\author[ad1]{Sinuo Zhang}
		\author[ad1]{Ivan Caicedo}
		\author[ad1,ad2]{Tomasz Hemperek}
		\author[ad1,ad3]{Toko Hirono}
		\author[ad1]{Jochen Dingfelder}
		
		\address[ad1]{Physikalisches Institut, Rheinische Friedrich-Wilhelms-Universit{\"a}t Bonn, Nu{\ss}allee 12, 53115 Bonn, Germany}
		\address[ad2]{Now at: DECTRIS AG, T{\"a}fernweg 1, Baden-D{\"a}ttwil, Switzerland}
		\address[ad3]{Now at: Institut f{\"u}r Prozess­daten­verarbeitung und Elektronik, Karlsruher Institut f{\"a}r Technologie,
Hermann-von-Helmholtz-Platz 1, 76344 Eggenstein-Leopoldshafen, Germany
}
		\begin{abstract}
		Silicon pixel sensors manufactured using commercial CMOS processes are promising instruments for high-energy particle physics experiments due to their high yield and proven radiation hardness. As one of the essential factors for the operation of detectors, the breakdown performance of pixel sensors constitutes the upper limit of the operating voltage. Six types of passive CMOS test structures were fabricated on high-resistivity wafers. Each of them features a combination of different inter-pixel designs and sets of floating guard rings, which differ from each other in the geometrical layout, implantation type, and overhang structure. A comparative study based on leakage current measurements in the sensor substrate of unirradiated samples was carried out to identify correlations between guard ring designs and breakdown voltages. TCAD simulations using the design parameters of the test structures were performed to discuss the observations and, together with the measurements, ultimately provide design features targeting higher breakdown voltages.
		\end{abstract}
		
	
	\end{frontmatter}


\section{Introduction}
\label{S1}
The operation of silicon detectors in high--energy physics (HEP) experiments usually requires a high reverse bias voltage on the sensors to provide a sufficiently high electric field and a possibly large depletion region to achieve fast charge collection by drift and a high signal-to-noise ratio.
The breakdown of the p--n junction emerges when the local electric field is able to trigger the impact ionisation and avalanche multiplication of charge carriers. With the consequence of a significantly high leakage current or even irreversible damage, such an effect is a limiting factor of the sensor operating voltage. 
Multiple floating guard rings are commonly used in power electronics devices
and silicon particle detectors for HEP experiments, to improve the breakdown
performance. By inserting ring-shaped implants surrounding the sensing area of
the sensor, the potential drop between the pixel matrix and the edge of the
chip can be smoothened and leads to a reduced electric
field~\cite{Baliga:2019vi}. Another purpose of the guard ring is to provide a
stable performance of the devices~\cite{BISCHOFF199327}.
The design and optimisation of guard rings have been an essential topic in
sensor design, from the earlier n-on-n sensors (e.g.~in~\cite{BISCHOFF199327,AVSET1996397,672633,rossi2006pixel,2009}), to the
n-on-p sensors (e.g.~in~\cite{BORTOLETTO1999178,EGOROV1999197,5603386,GALLRAPP201229}). The
steepness of the potential distribution and the subsequent strength of the
electric field can be manipulated by the design and fabrication features in the
guard ring region, such as the doping profile, the geometric design of the
rings, the oxide charge, and the oxide thickness~\cite{5603386}. The guard ring
design studied in~\cite{5603386} consists of p-type ring implants, where the
width of and spacing between implants are adjusted in a way that the
potential drops almost uniformly between them. The variation of adding
n-type ring between the p-type rings, as well as the overhang (field-plate)
structure at the guard rings were also studied
in~\cite{5603386,BORTOLETTO1999178}. In~\cite{GALLRAPP201229}, the studied
sensors were equipped with 15 and 19 floating guard rings, and the region
between the pixel matrix and the dicing edge covers a large area
(\SI{1040}{\micro\meter}).  \\
\indent
Commercial CMOS (Complementary Metal-Oxide-Se-miconductor) sensor manufacturing technologies offer sensor production with high yield, high throughput, and comparatively low cost on large wafers. 
Furthermore, the accessibility of features such as multi-metal layers,
polysilicon layers, and metal--insulator--metal (MIM) capacitors offers the
possibility to implement special sensor
characteristics~\cite{Pohl_2017,DIETER2021165771}. Several sensors produced in CMOS
technologies have proven to be radiation tolerant up to the requirements of HEP
experiments~\cite{Pohl_2017,DIETER2021165771,HIRONO201987,Barbero_2020}. Such
features make the sensor fabricated in commercial CMOS technologies a
promising instrument in the HEP experiments to meet the increasing demand for
silicon detectors~\cite{Garcia_Sciveres_2018}. However, challenges for
designing and optimising the guard ring structure of passive-CMOS sensors arise
from the predefined processing technology. Features like doping profiles and
the oxide properties are typically restricted to the commercial fabrication
line.\\
\indent
A comparative study of the impact of guard ring layout, implantation type, the
overhang of the guard ring implants, and the potential at the field-plate on
the breakdown voltage is presented in this paper. The guard ring designs
feature less number of rings and a smaller total width of the guard ring region
(\SI{274}{\micro\meter}), in comparison with previous designs in the
aforementioned literatures. Test structures fabricated in
LFoundry~\cite{lfoundry} $150~\mathrm{nm}$ CMOS process were investigated via leakage
current measurements, and their behaviour was qualitatively compared with
simulations using the Synopsys Technology Computer-Aided Design
(TCAD)~\cite{tcad}.

\section{The Test Structures}
Six different n-in-p passive CMOS test structures were designed and fabricated
in 2016 as a part of a multi-project wafer submission. The sensor structure,
consisting of p-type and n-type wells, was formed by the implantation processes
which are standardly utilised in high-voltage devices. A Czochralski-grown
p-type wafer with a foundry-specified typical resistivity of
$4 \sim 5~{\mathrm{k} \Omega}~\mathrm{cm}$ \cite{Pohl_2017} was used as substrate. Moreover, the chips were available in two thicknesses: $725~\si{\micro\meter}$ without backside processing and $200~\si{\micro\meter}$ with backside p-type doping and a metallisation layer. 

\subsection{Design}
\label{design}
\begin{figure}[t]
	\centering
	\includegraphics[width=0.5\textwidth]{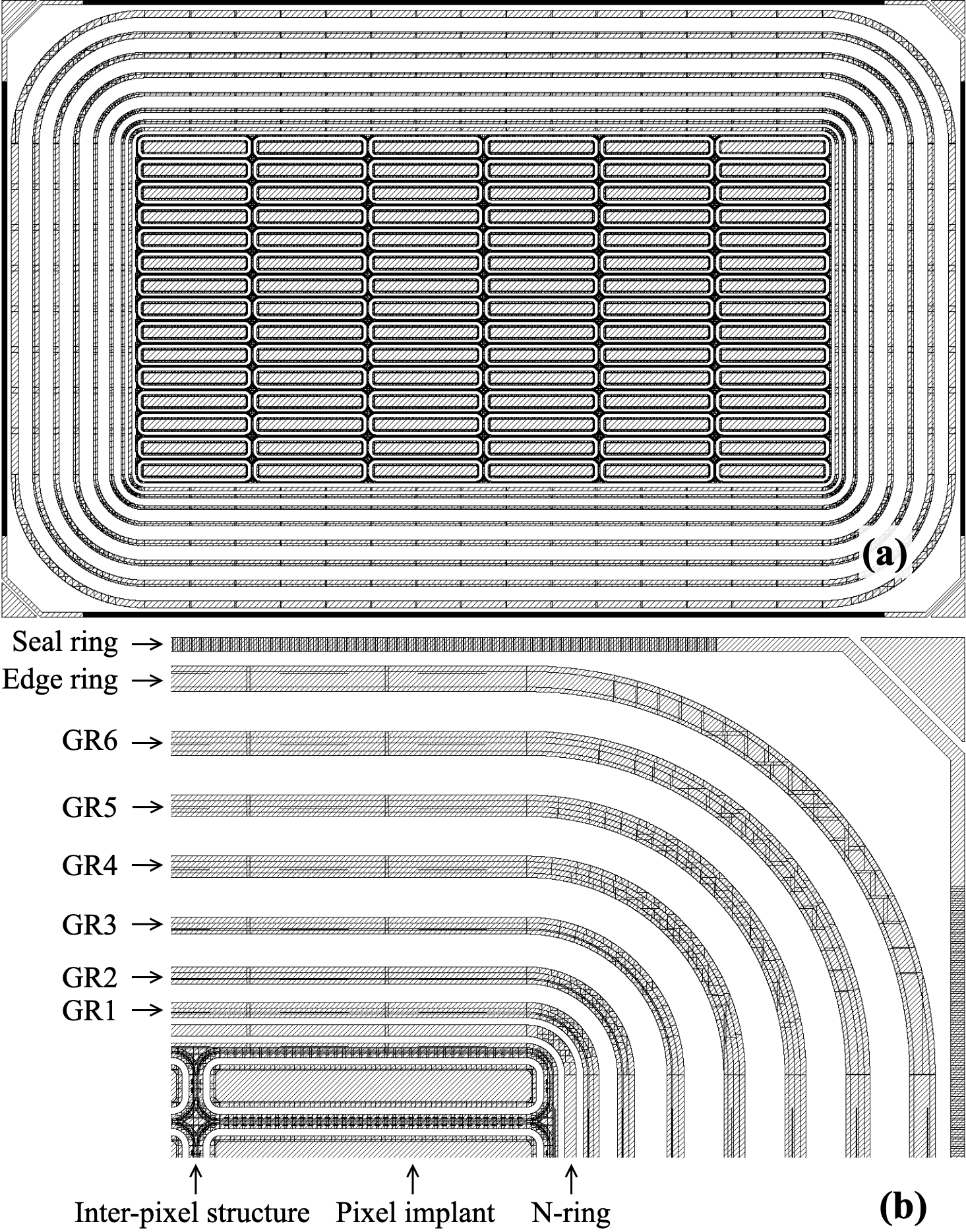}
	\caption{Generic layout of the test structures. The shaded areas represent the implants. (a): full layout of a test structure with the size of $\mathrm{\SI{2083}{\micro\meter}\times\SI{1334}{\micro\meter}}$. (b): The upper right corner of the test structure, shows the details of the pixel matrix region and the floating guard rings (GR). The edge of the test structure is indicated by the seal ring, which has a structure that protects the enclosed circuits, and ensures the product reliability of CMOS integrated circuits \cite{CHEN20051311}. The edge ring is electrically connected to the seal ring via metal wiring. All the listed structures can be individually accessed by lab instruments through metal pads (not illustrated).}
	\label{fullStrB}
\end{figure}

\begin{figure}[t]
	\includegraphics[width=0.45\textwidth]{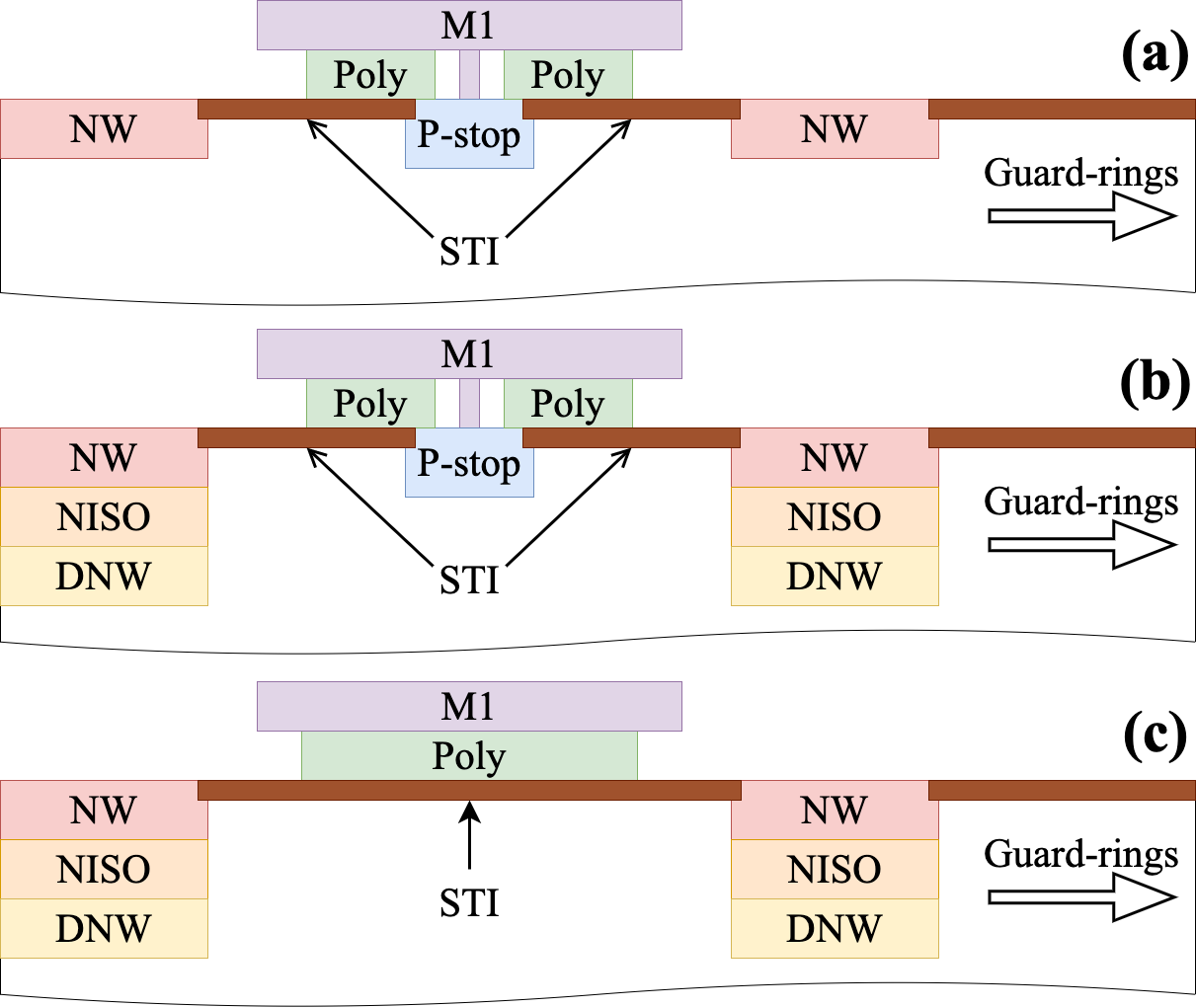}
	\caption{Schematic cross-section of the edge of the pixel matrix region. M1: the first metal layer in the CMOS process. (a): the single n-well as the implants for pixels and n-ring, and a p-stop with overhang as the inter-pixel structure (used in structure ``A"). (b): deep n-wells for pixels and n-ring, and a p-stop with overhang as the inter-pixel structure (used in ``B'', ``C'', ``D'', ``F''). (c): deep n-wells for pixels and n-ring, and a polysilicon field-plate as the inter-pixel structure (used in ``E'').}
	\label{InterPixel}
\end{figure}

A pixel matrix surrounded by multiple ring structures forms the generic layout of the test structures, as depicted in Figure \ref{fullStrB}(a). The pixel matrix region consists of $15~\times~6$ pixel-electrode implants (``Pixel implant''), the inter-pixel structures, and an n-type ring implant (``N-ring'') enclosing the entire matrix structure (Figure \ref{fullStrB}(b)). Concentric guard rings (GR) are located between the n-ring and the edge of the test structure. Six different test structures (labelled with letters ``A'' to ``E'', as summarised in Table \ref{TSprop}) differ in the number (spacing) of the guard rings, the type of doping used at the guard rings, and the structure in the inter-pixel region.\\
\begin{table*}[t]
\centering
\begin{tabular}{cccccc}
\hline
\hline
\textbf{Label} & \textbf{Implant} & \textbf{Overhang} & \textbf{\# Rings (GR1 Gap)} & \textbf{Inter-pixel structure} & \textbf{Deep n-well}\\
\hline
A & p 	& Yes 	& 6 ($\SI{8}{\micro\meter}$) & p-stop 		& No 	\\
B & n+p & Yes 	& 6 ($\SI{8}{\micro\meter}$) & p-stop 		& Yes 	\\
C & n+p & Yes 	& 5 ($\SI{32}{\micro\meter}$) & p-stop 		& Yes 	\\
D & p 	& Yes 	& 6 ($\SI{8}{\micro\meter}$) & p-stop 		& Yes 	\\
E & n+p & Yes 	& 5 ($\SI{32}{\micro\meter}$) & field-plate 	& Yes 	\\
F & n+p & No	& 5 ($\SI{32}{\micro\meter}$) & p-stop 		& Yes 	\\
\hline
\hline
\end{tabular}
\caption{Features of guard rings in the test structures. The corresponding descriptions can be found in the text and figures: Figure \ref{InterPixel} for the ``Inter-pixel structure'' and ``Deep n-well'', Figure \ref{GRDetail} for the `` \#Rings(GR1 Gap)'', and Figure \ref{GRFP} for the ``Implant'' and ``Overhang''.}
\label{TSprop}
\end{table*}
\indent
In the pixel matrix region, the test structures differ in the implantation profile of the pixel implants and the inter-pixel structure, as depicted in  Figure \ref{InterPixel}. The CMOS process offers three n-type well implantation profiles with increasing depth: standard n-well ``NW'' for MOSFET (Metal--Oxide--Semiconductor Field-Effect Transistor) fabrication, the ``NISO'', and the ``DNW''. These wells are employed to form the pixel implants and the n-ring in two configurations: a standard n-well or a deep n-well combining NW, NISO, and DNW. Moreover, two approaches of the inter-pixel structure were implemented: either a p-well (as in a standard ``p-stop'') with two overhang polysilicon (``Poly'') layers that exceed its edges, or a single polysilicon layer (``field-plate'') directly above the field oxide Shallow Trench Isolation (``STI'') forming a MOS (Metal--Oxide--Semiconductor) structure.  \\
\begin{figure}[h]
	\centering
	\includegraphics[width=0.45\textwidth]{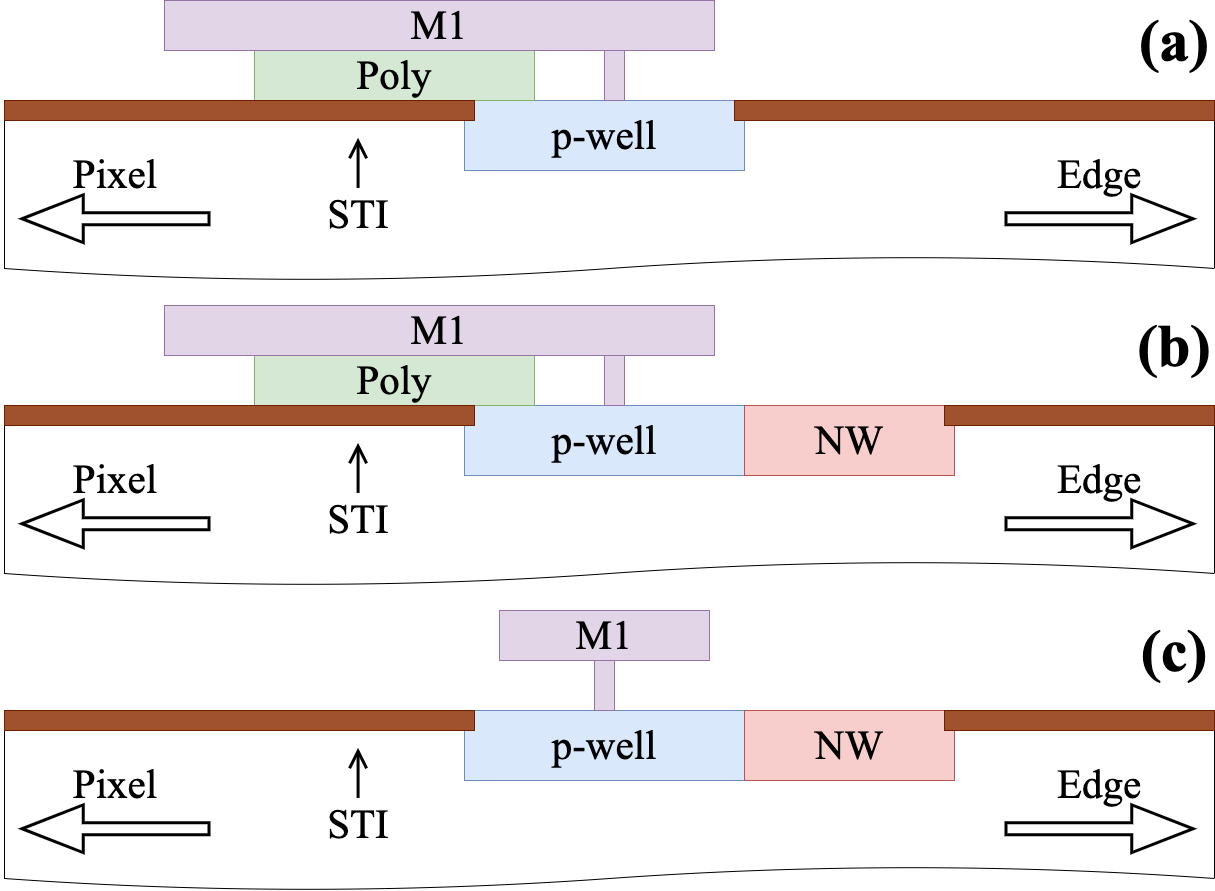}
	\caption{The profile of GR and the overhang structure. (a): p-well implant with polysilicon overhang for structure A and D; (b): n+p implant with polysilicon overhang for structure B, C, and E; (c): n+p implant without polysilicon overhang for structure F.}
	\label{GRFP}
\end{figure}
\begin{figure*}[h]
\centering
\includegraphics[width=0.8\textwidth]{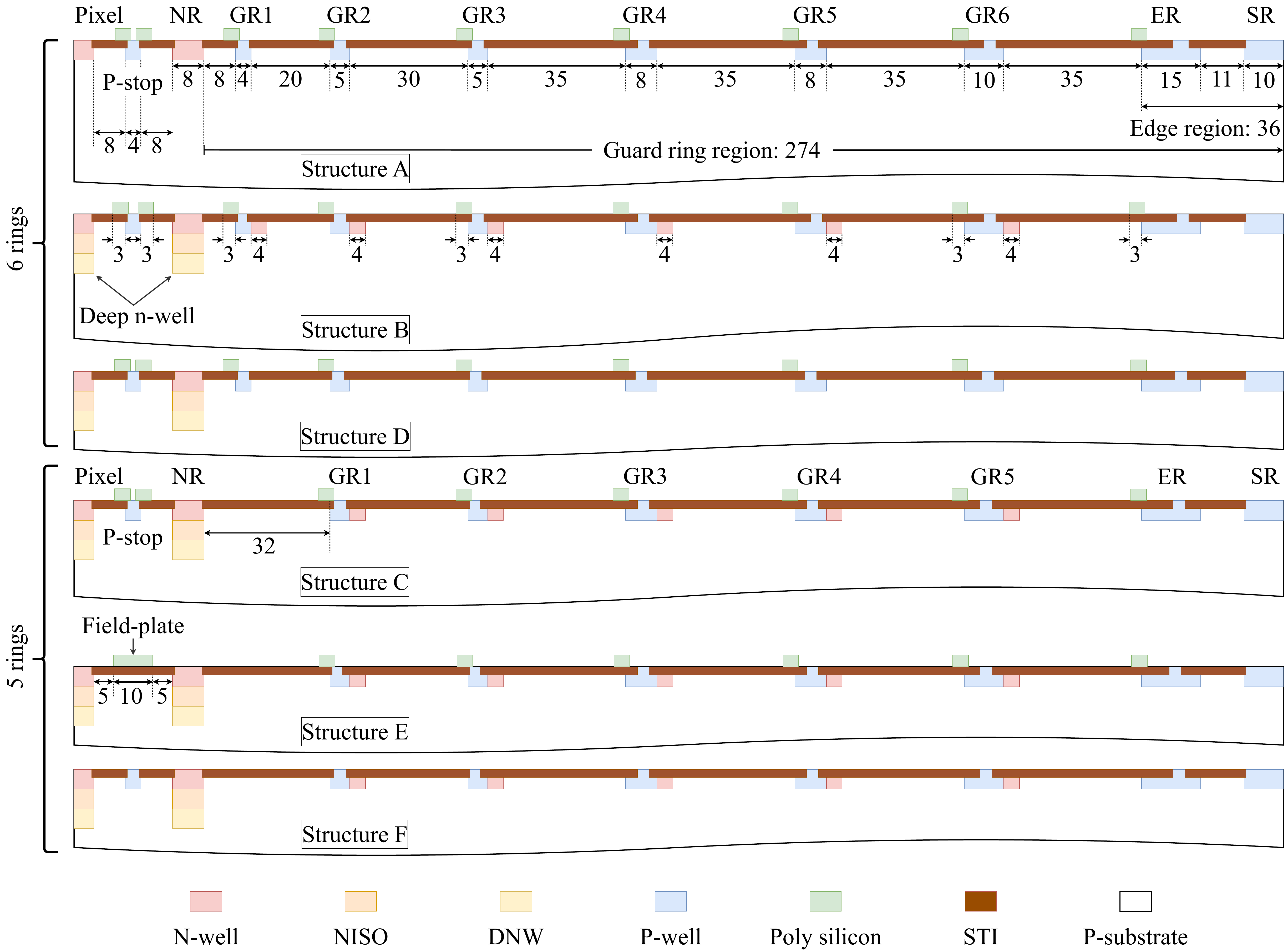}
\caption{Section views of the guard ring structures, consisting of a part of the pixel matrix (the ``Pixel'', ``P-stop''/``Field-plate'', and the n-ring ``NR''), the floating guard rings (``GR''), and the edge region (the edge ring ``ER''and the seal ring ``SR''). The dicing edge of samples is close to the seal ring. The designs are based on the geometry of ``structure A'', which was inspired by the design presented in \cite{5603386}, and it has 6 floating p-type guard rings (``GR1'' to ``GR6''). For the overhang and field-plate structures, only poly silicon layer is depicted for simplicity. The geometrical parameters are given in \SI{}{\micro\meter}, and the un-labeled parts have the same value as the given ones.}
\label{GRDetail}
\end{figure*}
\indent
When it comes to the guard rings, the test structures implement multiple designs that differ in the spacing between rings, the type of implant, and the use of polysilicon as an overhang. Figure \ref{GRFP} depicts three configurations for the guard-ring's implant and overhang. The guard rings are formed using the same implantation process as for the pixels, where the p-type ring is formed using the same p-well as for the p-stop. The second version of the guard ring implant is the so-called ``n+p'' type, where an n-well is attached to the outer side of the original p-type ring. By closely placing the p- and n-rings, the width of the guard ring region is reduced, with respect to the designs in \cite{BORTOLETTO1999178,5603386}. The overhang refers to the structures that use the conductive polysilicon and metal layers above the STI to form a MOS structure in the proximity of the guard ring implant. Its effects have been discussed in \cite{AVSET1996397,rossi2006pixel} for silicon sensors with an n-type substrate and in \cite{Hemperek:2018ut} for p-type substrate CMOS detectors, showing the ability to modify the local potential and electric field strength at the surface of the silicon substrate. Each configuration is applied to all guard rings of the corresponding test structure. Figure \ref{GRDetail} depicts the cross-section of the guard ring layouts, where the version consisting of six guard rings (structure A, B, and D) serves as a reference. The version with five guard rings (structure C, F, and E) is realised by removing the innermost guard ring, which effectively increases the spacing between the n-ring and the GR1 (GR1 gap). The edge ring is electrically connected with the seal ring in the design.  

\subsection{Samples, Measurement setup and Procedure}
Originally, each set of test structures A to E was part of a single die, and all of the test structures were located along the dicing edges. Two samples (TS1, TS2) with a thickness of $725~\si{\micro\meter}$ were prepared, where TS1 was fully diced so that each test structure was separated from the original die. All test structures with a thickness of $200~\si{\micro\meter}$ came from seven different samples (S1--S7) which were not fully diced to be measured. 
\begin{figure}[t]
	\centering
	\includegraphics[width=0.45\textwidth]{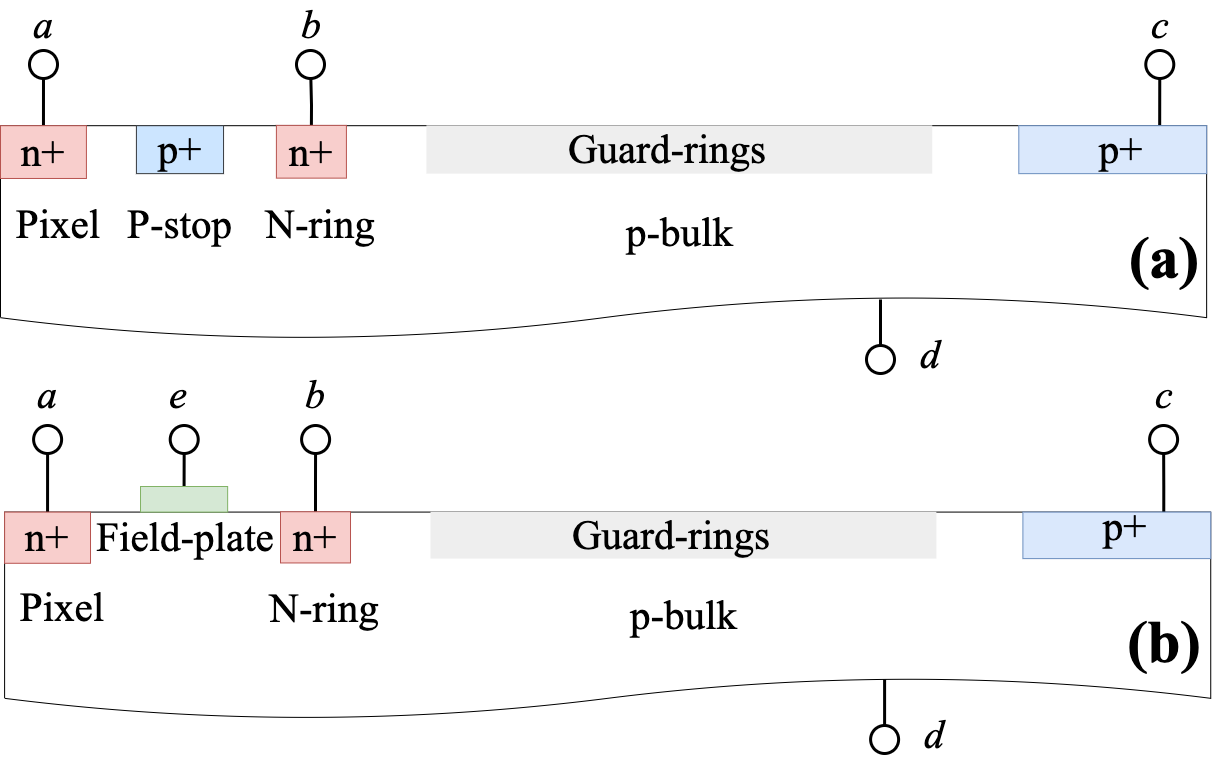}
	\caption{Contact points ``\textit{a}''-``\textit{e}'' to apply voltages and/or perform measurements in structures (a): A, B, C, D, F, and (b): E.  Floating guard rings are not explicitly depicted. Contact ``\textit{c}" is used for the frontside biasing, ``\textit{d}" to bias a backside processed sensor, and ``\textit{e}" to apply a desired voltage to the inter-pixel field-plate in structure E. The p-stop is floating in all the cases.}
	\label{meas}
\end{figure}

Breakdown voltages $V_{\mathrm{BD}}$ were determined by identifying the voltage at
which a sharp increase in current occurs while scanning the current--voltage
behaviour (``I-V curve'') of reverse biased sensors~\cite{sze2007physics}. Figure \ref{meas} illustrates multiple points where measurements can be performed or voltages can be set to the implants in every test structure.
All $725~\si{\micro\meter}$ and $200~\si{\micro\meter}$ thick samples can be biased from the front side
of the sensor~\cite{Baselga_2018} by applying a voltage at contact
``\textit{c}'' and ground the pixel at contact ``\textit{a}''. Alternatively,
the latter ones can be biased from their backside through contact
``\textit{d}'', since they have been processed and metallised after thinning. 
In any of these cases, the current at the pixel matrix can be measured at contact ``\textit{a}'' for two scenarios: (1) a floating n-ring or (2) a grounded n-ring with contact \textit{b} connected with \textit{a}. The grounded n-ring scenario represents the designed operation condition of such pixel sensors, where a potential drop is formed from the n-ring across the guard rings to the edge after applying the bias voltage.
In the floating n-ring scenario, the n-ring and the p-stop (or the field-plate),
will become a part of the entire potential drop from the pixel to the edge.
Although the p-stop or the field-plate is originally an inter-pixel isolation
structure, the floating n-ring scenario makes it act as an effective floating
guard ring, and it helps to understand the effect from the deep n-well and the
field-plate on the breakdown performance. For structure E, an additional
voltage can be applied at the polysilicon field-plate via contact
``\textit{e}''. Due to the MOS structure, such a voltage has influence on the
potential at the silicon surface beneath. The study in~\cite{1474639} used a
similar structure and measurement setting showing the change in the breakdown
voltage, but the field-plate was placed at the edge of an implant to modify the
local potential distribution. This result inspired the application of the
overhang structure~\cite{Baliga:2019vi}. Here, the field-plate in structure E
is used to study its influence on the overall potential distribution. \\
\indent
In this work, the leakage current of all test structures was measured while they were frontside-biased. The $725~\si{\micro\meter}$ thick samples were examined with floating n-ring, whereas on the $200~\si{\micro\meter}$ thick structures both the floating and grounded n-ring cases were measured. For structure E, different potentials were applied at the field-plate. The ramping of bias voltage was done in steps of \SI{3}{\volt} within \SI{5}{\second} for the $725~\si{\micro\meter}$ thick samples. For $200~\si{\micro\meter}$ thick samples, a step of \SI{5}{\volt} within \SI{5}{\second} was used. For each step the average over 5 measurements was taken, and both the standard deviation and the step size were considered for error estimation.\\

\section{TCAD Simulation for a Qualitative Study}
A 2D cross-section of the guard ring region, as depicted in Figure \ref{GRDetail},
was employed as the simulation domain. This was created by simulating the
manufacturing process using the tool ``Sentaurus
Process''~\cite{sentaurusProcess} within the framework of Synopsys TCAD. The
resistivity of the sensor bulk was set to $5~{\mathrm{k} \Omega}~\mathrm{cm}$, representing a boron
doping concentration of approximately $2.7\times 10^{12}~{\mathrm{cm}^{-3}}$, which is within the
foundry-specified range. Breakdown simulations of the given guard ring
structures were performed using the device simulation tool ``Sentaurus
Device''~\cite{sentaurusDevice}. Charge carrier mobility models, such as
Masetti doping dependency~\cite{1483108}, Canali high-field
saturation~\cite{1478102} and Conwell--Weisskopf carrier--carrier
scattering~\cite{PhysRev.77.388} were taken into account to describe the
complex conditions in the semiconductor sensor. The van Overstraeten impact
ionisation model~\cite{Van_Overstraeten_1970} with the parameters given
in~\cite{sentaurusDevice} was employed to simulate the avalanche breakdown
effect. The Shockley--Read--Hall recombination
model~\cite{Shockley:1952aa,PhysRev.87.387} with Hurkx tunnelling
model~\cite{121690} was adopted to include the effect of impurities at
${\mathrm{Si/SiO}_{2}}$ interface for unirradiated sensors~\cite{Morozzi_2021}. 
For a qualitative study, a set of dummy parameters for oxide charge
concentration ($Q_{\mathrm{OX}}=1.0\times10^{10}\,{\mathrm{cm}^{-2}}$) and interface trap concentrations ($N_{\mathrm{acceptor}}=1.0\times10^{9}\,{\mathrm{cm}^{-2}}$ and
$N_{\mathrm{donor}}=1.0\times10^{9}\,{\mathrm{cm}^{-2}}$) based on~\cite{Morozzi_2021} was applied to the simulations, since
the quantitative characteristics of these test structures are unknown.

\section{Measurement Results}
\subsection{$725~\si{\micro\meter}$ Thick Structures with Floating N-ring}
\label{BDIV700NW}

\begin{figure}[t]
	\centering
	\includegraphics[width=0.48\textwidth]{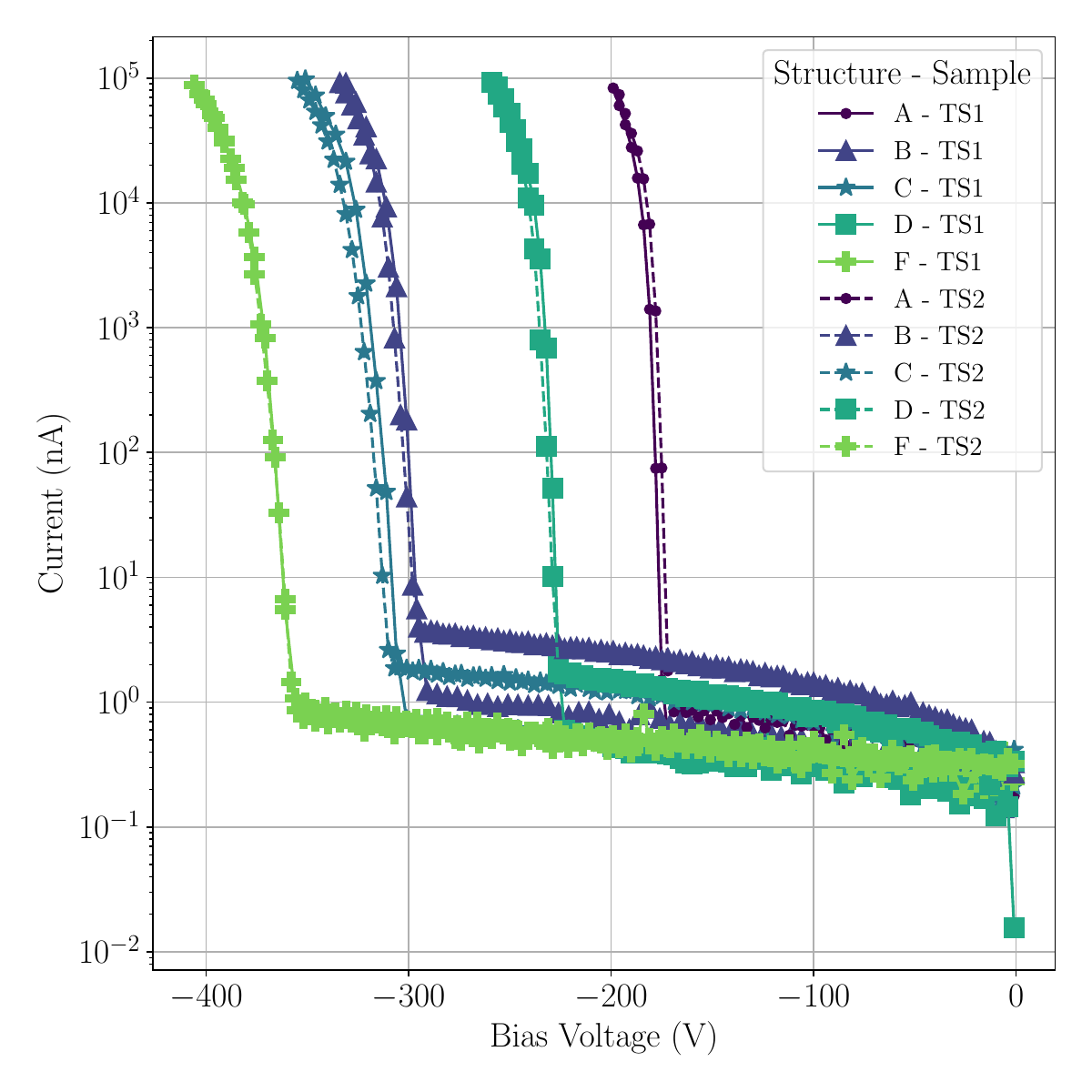}
	\caption{I-V Curves for two sets of $\SI{725}{\micro\meter}$ test structures with a grounded n-well and floating n-ring. The fully diced samples (TS1) reveal the same breakdown voltage as the not-diced samples. The relation between the breakdown voltages of guard ring test structures: $V_\mathrm{BD,F}>V_\mathrm{BD,C}>V_\mathrm{BD,B}>V_\mathrm{BD,D}>V_\mathrm{BD,A}$.}
	\label{700FNR}
\end{figure}

The I-V curves of two different samples of structures A, B, C, D, and F are presented in Figure \ref{700FNR} and the extracted breakdown voltages are listed in 
Table \ref{700BDfNR}. Both samples provided consistent results, where no influence from the dicing process was observed. The breakdown voltages of the structures ranged from $\SI{-175}{V}$ (structure A) to $\SI{-361}{V}$ (structure F), indicating a dependence on the guard ring design.\\
\begin{table}[b]
\centering
\begin{tabular}{c|c||c|ccc}
\hline
\hline
\textbf{Label} & \textbf{$V_\mathrm{BD}$ (V)} & \textbf{Label}& \textbf{$V_\mathrm{poly}$ (V)} & \textbf{$V_\mathrm{BD}$ (V)}\\
\hline
A & $-175 \pm 3$ & & $0$	& $-361 \pm 3$	\\
B & $-295 \pm 3$ & & $-50$	& $-415 \pm 3$	\\
C & $-310 \pm 3$ &E& $-100$ & $-475 \pm 3$  \\
D & $-226 \pm 3$ & & $-150$	& $-190 \pm 3$	\\
F & $-361 \pm 3$ & & $=V_\mathrm{bias}$	& $-166 \pm 3$	\\
\hline
\hline
\end{tabular}
\caption{Breakdown voltages of $725~\si{\micro\meter}$ test structures with floating n-ring. The $V_\mathrm{BD}$ and the corresponding errors are extracted from the samples presented in Figures \ref{700FNR} and \ref{700StrE}.}
\label{700BDfNR}
\end{table}
\begin{figure}[h]
	\centering
	\includegraphics[width=0.48\textwidth]{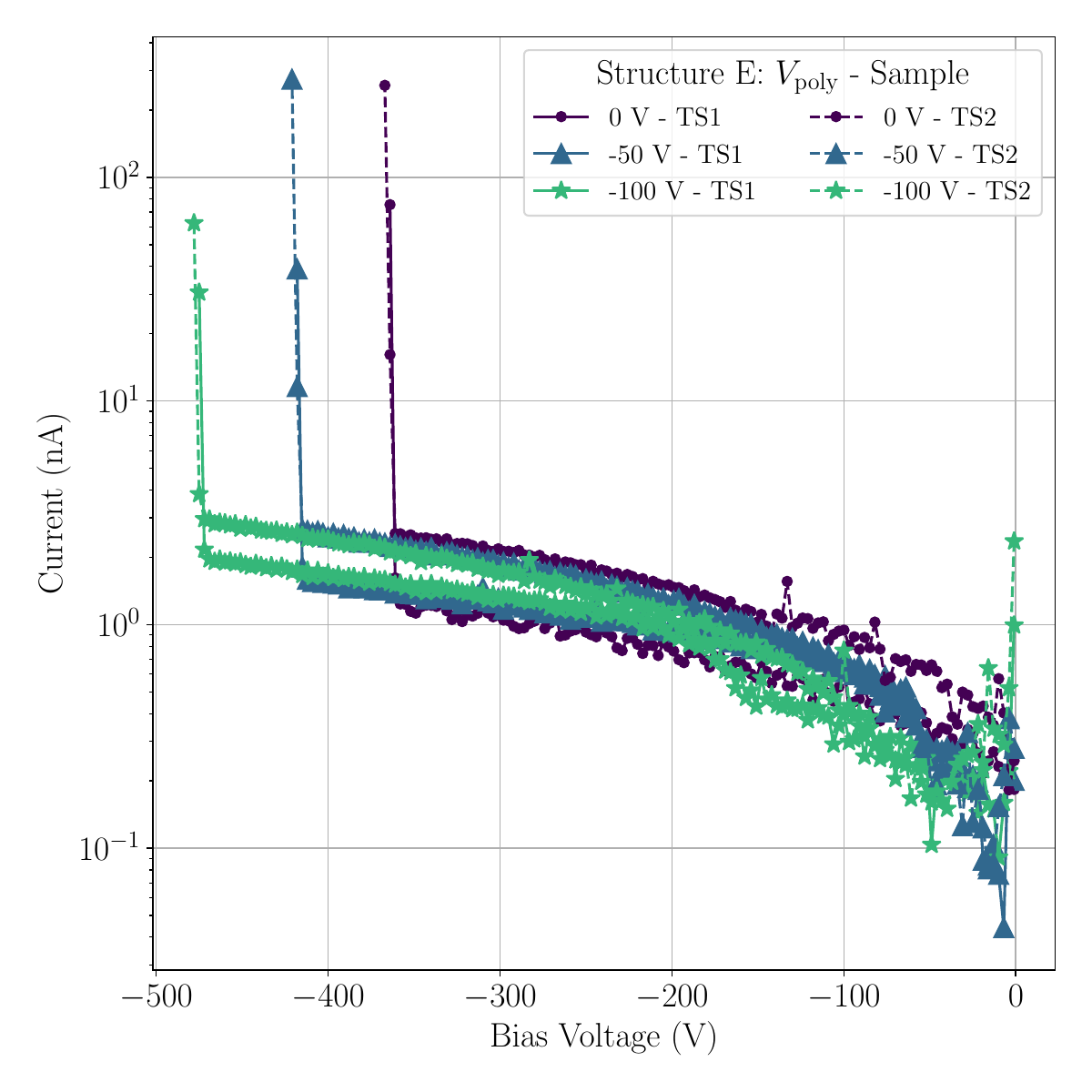}
	\caption{I-V curves for two samples of test structure ``E'' (725um thickness). The change of the poly silicon voltage results in the change of the same amount in the breakdown voltage, i.e. $\Delta V_\mathrm{poly}=\Delta V_\mathrm{BD}$.}
	\label{700StrE}
\end{figure}
\indent
The results of the structure E with fixed potentials at the polysilicon field-plate ($V_{\mathrm{poly}}$) are presented in Figure \ref{700StrE} and Table \ref{700BDfNR}. The onset of the breakdown effect was delayed by the
increased negative potential between $0~\mathrm{V}$ and $-100~\mathrm{V}$ at the
field-plate. This linear relation between $\vert\Delta V_{\mathrm{poly}}\vert$ and $\vert\Delta V_{\mathrm{BD}}\vert$ agrees
with the relation found in~\cite{1474639}. Moreover, connecting the field-plate
with the bias voltage ($V_{\mathrm{poly}}=V_{\mathrm{bias}}$) led to an earlier breakdown at $V_{\mathrm{bias}}=-166~\mathrm{V}$.
For $V_{\mathrm{poly}}=-\SI{150}{\volt}$, the breakdown voltage decreased to $-190~\mathrm{V}$, and the
original behaviour of the sample could not be recovered. A floating field-plate
resulted in non-reproducible measurement results.\\
\indent
The test structures were paired in four different groups to determine the impact of individual design features on the breakdown voltage, as presented in Table \ref{700BDfNRcomp}. A higher breakdown voltage was observed for the structures with deep n-implant at pixels and n-ring (Figure \ref{InterPixel}(b)), large spacing between n-ring and the first guard ring (the ``5 rings'' layouts in Figure \ref{GRDetail}), additional n-implant at the guard rings, and the absence of the polysilicon overhang at the guard rings (Figure \ref{GRFP}(c)). Structure E exhibits a possibility of increasing the breakdown voltage by applying an additional negative potential at the field-plate.

\subsection{\SI{200}{\micro\meter} Thick Structures with Floating N-ring}
\label{BDIV200NW}
The measurement of breakdown voltages for samples with $200~\si{\micro\meter}$ thickness delivered similar results as for the $725~\si{\micro\meter}$ sensors, as summarised in 
\ref{200BDfNR}. It is noticeable by comparing 
Tables \ref{200BDfNR} with \ref{700BDfNR} that larger variations of the breakdown voltage were observed for the $200~\si{\micro\meter}$ samples. This was likely due to variations between samples after back-side processing. \\
\indent
Measurements of the structure E with
$V_{\mathrm{poly}}=\{\SI{0}{\volt},\allowbreak~\SI{-50}{\volt},~\SI{-100}{\volt}\}$ on two samples have delivered the same relation between $V_{\mathrm{poly}}$ and $V_{\mathrm{BD}}$ (Figure 
\ref{200BDfNR}) as observed in the $725~\si{\micro\meter}$ thick samples. A discrepancy of approximately $50~\mathrm{V}$ in $V_{\mathrm{BD}}$ between both samples results in large variations of the breakdown voltages. Due to the consistent behaviour between the $V_{\mathrm{BD}}$ and $V_{\mathrm{poly}}$, as well as the similar leakage current level, this discrepancy is attributed to variations between samples after dicing and back-side processing. Comparing with the results from  \ref{BDIV700NW}, the consistent relation between guard ring designs (or $V_{\mathrm{poly}}$) and $V_{\mathrm{BD}}$ of structures are independent of the thickness of samples. Therefore, the comparison concerning the control groups in Table \ref{700BDfNRcomp} still applies.\\
\begin{table}[t]
\centering
\begin{tabular}{c|l||c|ccc}
\hline
\hline
\textbf{Label} & \textbf{$V_\mathrm{BD}$ (V)} & \textbf{Label}& \textbf{$V_\mathrm{poly}$ (V)} & \textbf{$V_\mathrm{BD}$ (V)}\\
\hline
A & $-174 \pm 3$ & & $0$	& $-400 \pm 25$	\\
B & $-308 \pm 5$ & & $-50$	& $-450 \pm 25$	\\
C & $-329 \pm 5$ &E& $-100$ & $-505 \pm 25$ \\
D & $-235 \pm 4$ & & $-150$ & $-208 \pm 6$	\\
F & $-367 \pm 18$ &	\\
\hline
\hline
\end{tabular}
\caption{Breakdown voltages of the $200~\si{\micro\meter}$ test structures with floating n-ring. Sample S1 and S4 were used for extracting $V_\mathrm{BD}$ of structure E. The same set of samples presented in Figure \ref{200GNR} are used for extracting $V_\mathrm{BD}$ of the rest of structures. }
\label{200BDfNR}
\end{table}
\begin{table*}[t]
\centering
\begin{tabular}{lll}
\hline
\hline
\textbf{Control Group} & \textbf{Compared Parameter} & \textbf{Condition for higher breakdown voltage}\\
\hline
A \& D & Deep implants in the active region	& Deep implants at the pixels and n-ring (D)\\
B \& C & Size of GR1 gap 	& Larger spacing between n-ring and GR1 (C)	\\
B \& D & N-implant at the guard rings 			& With n-implant at the guard rings (B)	\\
C \& F & polysilicon overhang	& Without polysilicon overhang structure (F)	\\
\hline
\hline
\end{tabular}
\caption{Comparison between structures with floating n-ring.}
\label{700BDfNRcomp}
\end{table*}

\subsection{$200~\si{\micro\meter}$ Thick Structures with Grounded N-ring}
\label{BDIV200GNR}
\begin{figure}[t]
	\centering
	\includegraphics[width=0.48\textwidth]{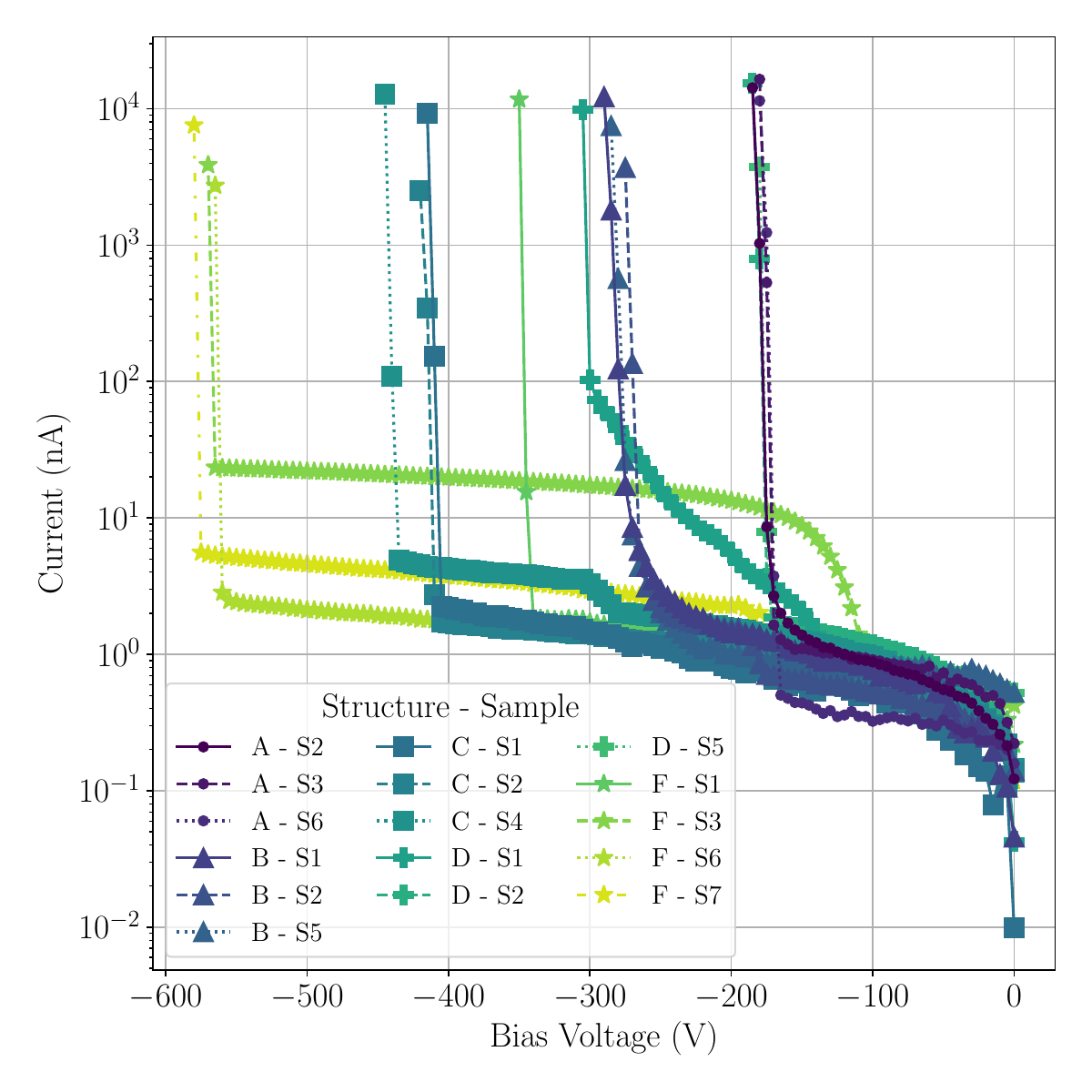}
	\caption{I-V Curves for various sets of test structures with a grounded n-ring. The relation between the breakdown voltages of guard ring test structures: $V_\mathrm{BD,F}>V_\mathrm{BD,C}>V_\mathrm{BD,B}>V_\mathrm{BD,D}> V_\mathrm{BD,A}$.}
	\label{200GNR}
\end{figure}

The same set of samples was tested with grounded n-ring for structures A, B, C, D, and F, as illustrated and summarised in Figure \ref{200GNR} and Table \ref{200BDGNR}. Both structures D and F on sample S1 showed a distinct breakdown performance, with respect to the other samples. These two specific measurements were neglected since they are likely caused by production errors unable to be confirmed by the measurements within this work. 
An abnormal increase of the leakage current of bias voltage above $\SI{100}{\volt}$
is observed for structure F from sample S3 and S7. The same effect has occurred
in the measurements with floating n-ring. This may be caused by imperfection of
the production or the backside processing of the
wafer~\cite{Pohl_2017,Hirono:2019aa}. Nevertheless, the current increase tends
to saturate at a certain level, and the breakdown voltage is in accordance with
the other samples.\\
\begin{table}[t]
\centering
\begin{tabular}{c|l||c|ccc}
\hline
\hline
\textbf{Label} & \textbf{$V_\mathrm{BD}$ (V)} & \textbf{Label}& \textbf{$V_\mathrm{poly}$ (V)} & \textbf{$V_\mathrm{BD}$ (V)}\\
\hline
A & $-175 \pm 4$ 	& 	& $0$	& $-395 \pm 30$	  \\
B & $-279 \pm 5$ 	& 	& $-50$	& $-385 \pm 25$	  \\
C & $-424 \pm 13$ 	& E & $-100$	& $-373 \pm 23$	 \\
D & $-180 \pm 4$ 	& 	& $-150$	& $-365 \pm 21$	  \\ 
F & $-569 \pm 7$ 	&  \\
\hline
\hline
\end{tabular}
\caption{Breakdown voltages of the $200~\si{\micro\meter}$ test structures with grounded n-ring. Sample S1 and S4 were used for extracting $V_\mathrm{BD}$ of structure E. The samples presented in Figure \ref{200GNR} are used for extracting $V_\mathrm{BD}$ of the rest of structures.}
\label{200BDGNR}
\end{table}
\begin{table*}[h]
\centering
\begin{tabular}{lll}
\hline
\hline
\textbf{Control Group} & \textbf{Compared Parameter} & \textbf{Condition for higher breakdown voltage}\\
\hline
A \& D & Deep implants in the active region	& None (--)\\
B \& C & size of GR1 gap 	& Larger spacing between n-ring and GR1 (C)	\\
B \& D & N-implant at the guard rings 			& With n-implant at the guard rings (B)	\\
C \& F & polysilicon overhang	& Without polysilicon overhang structure (F)	\\
\hline
\hline
\end{tabular}
\caption{Comparison between $200~\si{\micro\meter}$ test structures with grounded n-ring.}
\label{200BDGNRcomp}
\end{table*}
\indent
With the exception of test structure A, the breakdown voltages measured for all samples with grounded n-ring (Table \ref{200BDGNR}) varied with respect to the measurements with floating n-ring (Table \ref{700BDfNR}). Lower breakdown voltages are observed for structure B and D, revealing a reduction of approximately $30~\mathrm{V}$ and $50~\mathrm{V}$, respectively. On the contrary, structure C and F exhibit an increase of breakdown voltage by approximately $100~\mathrm{V}$ and $200~\mathrm{V}$, respectively. Nevertheless, the comparison in the control groups (
Table \ref{200BDGNRcomp}) mostly agrees with Table \ref{700BDfNRcomp}, except for structure D showing a similar breakdown performance as structure A.\\
\indent
The results in Table \ref{200BDGNR} indicate that for grounded n-ring the negative potential at the polysilicon field-plate has less influence on the breakdown performance of structure E, in comparison with the floating n-ring scenario. Adding $50~\mathrm{V}$ negative voltage at the field-plate, the decrease of the breakdown voltage is only approximately $10~\mathrm{V}$.. 

\subsection{Implementation of the Guard-ring Designs in Other Devices}
Guard ring designs similar to those used by the test structures of this work
were already implemented in large sensors ($\sim\mathcal{O}~({\mathrm{cm}^{2}})$) fabricated in the same
CMOS process. In the prototype of the depleted monolithic active pixel sensor
(DMAPS) LF-CPIX~\cite{HIRONO201987,Hirono:2019aa}, two guard ring layouts akin
to the one in structure D were employed. The difference between both of them
was the spacing between the n-ring and GR1, where one was smaller (with
$V_{\mathrm{BD}}\approx \SI{-130}{\volt}$) and the other one was larger (with $V_{\mathrm{BD}}\approx \SI{-220}{\volt}$) than the gap used
for the test structure. Another DMAPS device LF-Monopix1~\cite{Wang_2020}
adopted a guard ring design based on the structure D of this work, with a gap
size larger than that in LF-CPIX. The resulting breakdown voltage was between
$-\SI{260}{\volt}$ and
$-\SI{300}{\volt}$~\cite{Hirono:2019aa,Wang_2020,Wang_2018,IGUAZ2019652}. In this
case, the influence of the spacing on the breakdown performance was also
validated. The successor of the LF-Monopix1, the
LF-Monopix2~\cite{DINGFELDER2022166747} has a guard ring design close to that
of the test structure F, which visibly improved its breakdown voltage
($-\SI{460}{\volt}\sim - \SI{500}{\volt}$)~\cite{Caicedo:2023aa}. This result is consistent with the
breakdown measurements of the test structures, where the additional n-well
attached to the floating p-type guard rings improves the breakdown voltage.

\section{TCAD Simulation of Breakdown Performances}

\begin{figure}[t]
	\centering
	\includegraphics[width=0.48\textwidth]{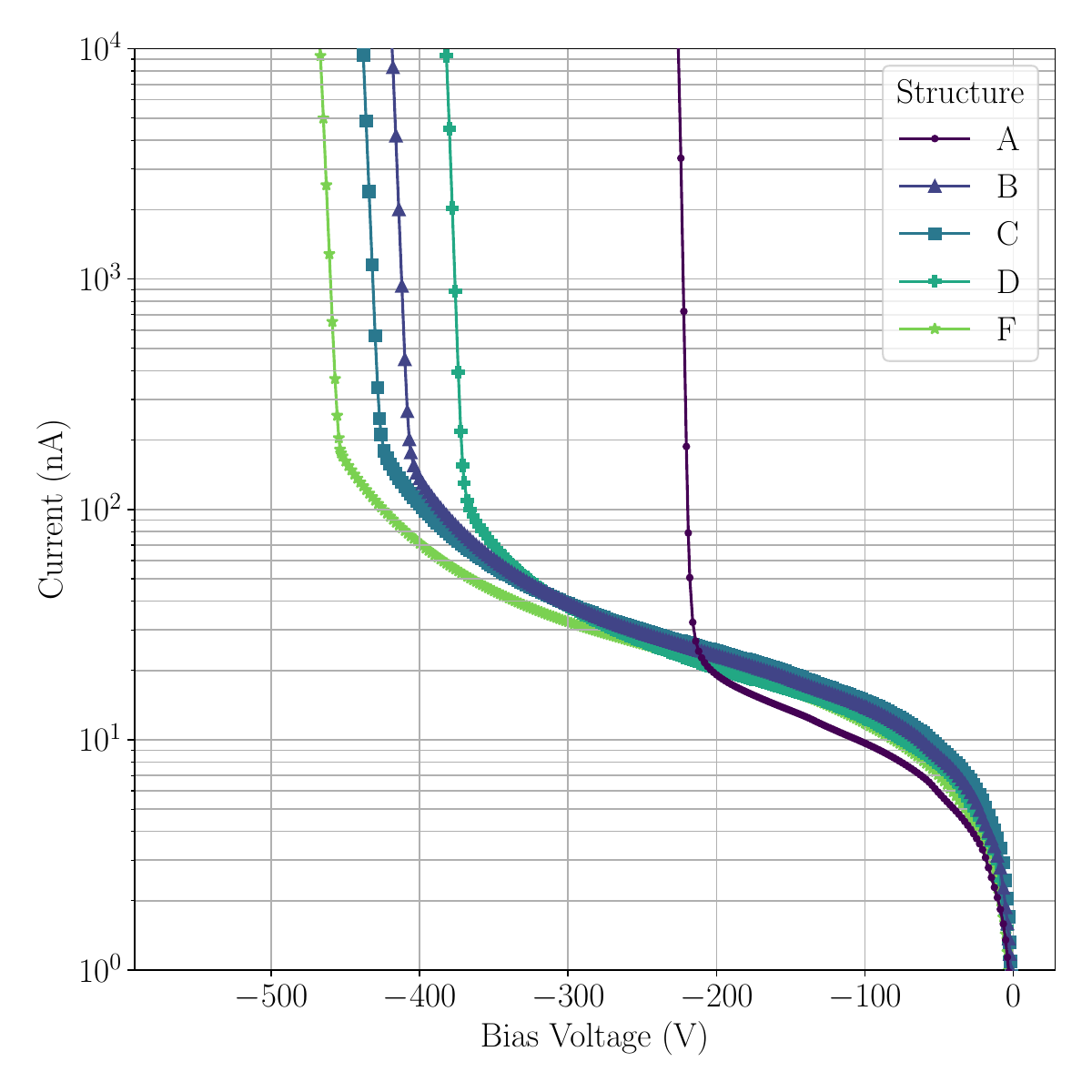}
	\caption{Simulated IV curves for structures A, B, C, D, and F with a floating n-ring. The relation between the breakdown voltages of guard ring test structures is the same as that obtained from Figure \ref{700FNR}.}
	\label{simBDIVfN}
\end{figure}

\begin{figure}[t]
	\centering
	\includegraphics[width=0.48\textwidth]{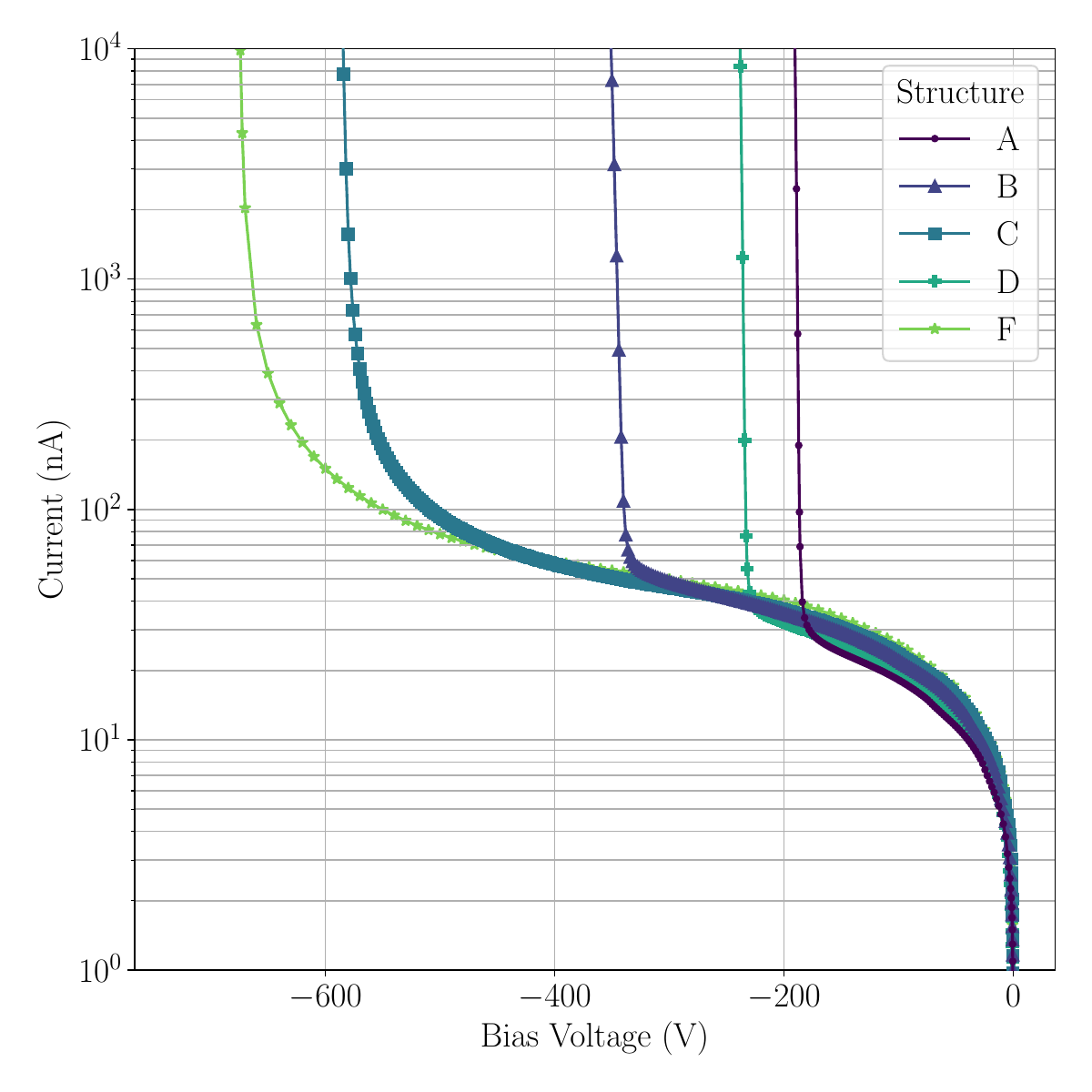}
	\caption{Simulated IV curves for structures A, B, C, D, and F with a grounded n-ring. The relation between the breakdown voltages of guard ring test structures is the same as that obtained from Figure \ref{200GNR}.}
	\label{simBDIVgN}
\end{figure}

\begin{figure*}[t]
	\centering
	\includegraphics[width=1\textwidth]{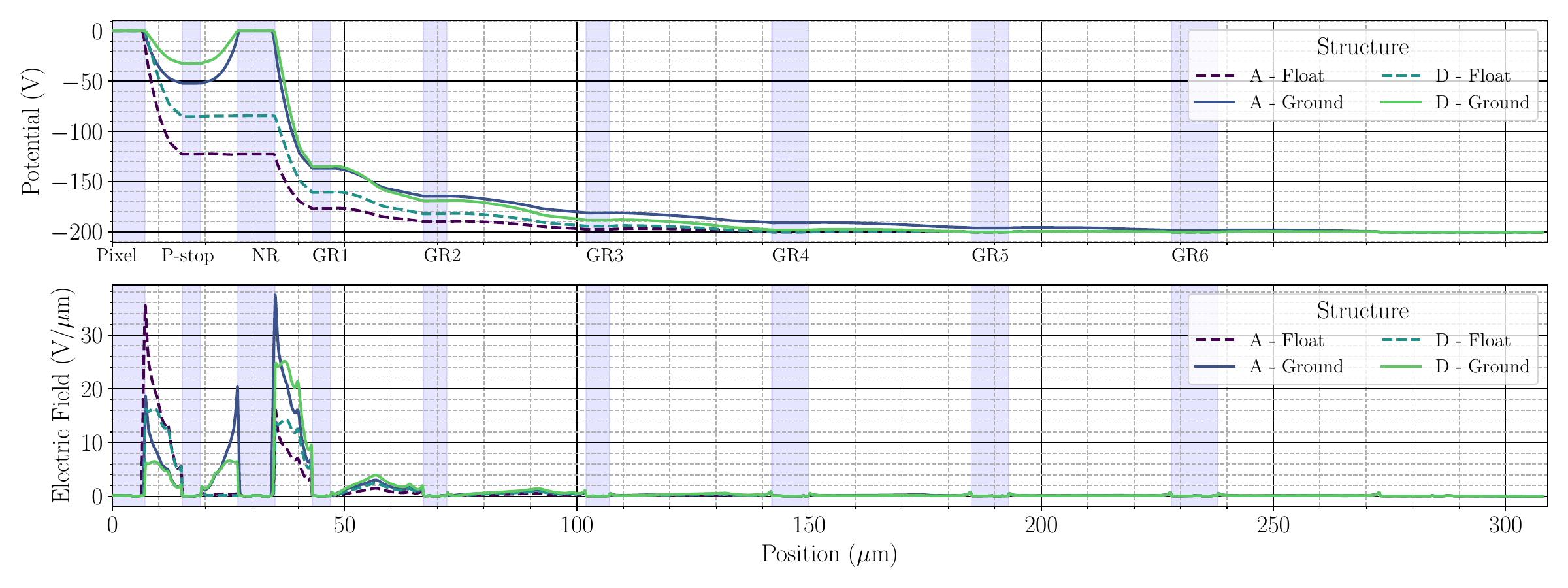}
	\caption{Potential (upper) and electric field strength (lower) distributions of structures A and D with $V_\mathrm{bias}=-\SI{200}{\volt}$. The highlighted pillar indicates the position and size of the wells, which are labelled according to Figure \ref{GRDetail}. The distributions indicated by the dashed lines represent the floating n-ring scenario, and the solid lines represent the grounded n-ring scenario. In the grounded n-ring scenario, the maximum electric field locates at the edge of the n-ring (NR), whereas this is shifted to the edge of the pixel when the n-ring is floating. The potential at the floating n-ring using a deep n-well implant results delivers a higher floating potential, which reduces the maximum electric field.}
	\label{simPotEfStrADgfN}
\end{figure*}

The simulated I-V curves of the test structures A, B, C, D, and F for floating and grounded n-ring are presented in Figures \ref{simBDIVfN} and \ref{simBDIVgN}, where the leakage currents are scaled according to the geometry of the test structure. Since this is merely a 2D simulation of the guard ring region and the actual processing technology may not be fully described by the simulation, the results only allow for a qualitative description of the behaviour observed in the actual samples. The extracted breakdown voltages are listed in Table \ref{simBDIV200}, and they reproduced the relative changes in breakdown voltage between structures for both floating or grounded n-ring cases. As observed in the measurements, the reduction of breakdown voltages for structure B (by $75~\mathrm{V}$) and D (by $158~\mathrm{V}$), and the increase of the breakdown voltages for structure C (by $148~\mathrm{V}$) and F (by $225~\mathrm{V}$) are obtained from the simulation.\\
\begin{table*}[t]
\centering
\begin{tabular}{c|ll||c|cll}
\hline
\hline
\textbf{Label} & \textbf{$V_\mathrm{BD,FN}$ (V)}& \textbf{$V_\mathrm{BD,GN}$ (V)} & \textbf{Label}& \textbf{$V_\mathrm{poly}$ (V)} & \textbf{$V_\mathrm{BD,FN}$ (V)} & \textbf{$V_\mathrm{BD,GN}$ (V)}\\
\hline
A & $-221$ & $-188$ & 	& $0$		& $-623$ &  $-600$	  	\\
B & $-409$ & $-344$ & 	& $-50$		& $-685$ &	$-583$  	\\
C & $-429$ & $-572$ & E & $-100$	& $-742$ &	$-572$ 		\\
D & $-374$ & $-235$ & 	& $-150$	& $-490$ &	$-569$  	\\ 
F & $-457$ & $-653$ &  	&			& 	 	 &				\\
\hline
\hline
\end{tabular}
\caption{Simulated breakdown voltages of the $200~\si{\micro\meter}$ test structures. The results for a floating and grounded n-ring are indicated by $V_\mathrm{BD,FN}$ and $V_\mathrm{BD,GN}$, respectively. The breakdown voltages are extracted as the voltage where the current exceeds \SI{1}{\micro\ampere}.}
\label{simBDIV200}
\end{table*}
\indent
To understand the relation between the breakdown performance and the design features of guard rings, it is helpful to consider the potential and electric field strength distribution across the surface of the guard ring region, because the strength of the impact ionisation and the onset of the junction breakdown are correlated with the magnitude of the electric field. The distributions were obtained at a depth of $100~\mathrm{nm}$ from the silicon/oxide interface (i.e.~the lower boundary of the STI). From these, the maximum electric field (or the largest potential-drop between implants) can be identified and compared across guard ring designs and measurement scenarios. Figure \ref{simPotEfStrADgfN} illustrates the extracted distributions for structures A and D with floating and grounded n-ring.  

\begin{figure}[H]
	\centering
	\includegraphics[width=0.48\textwidth]{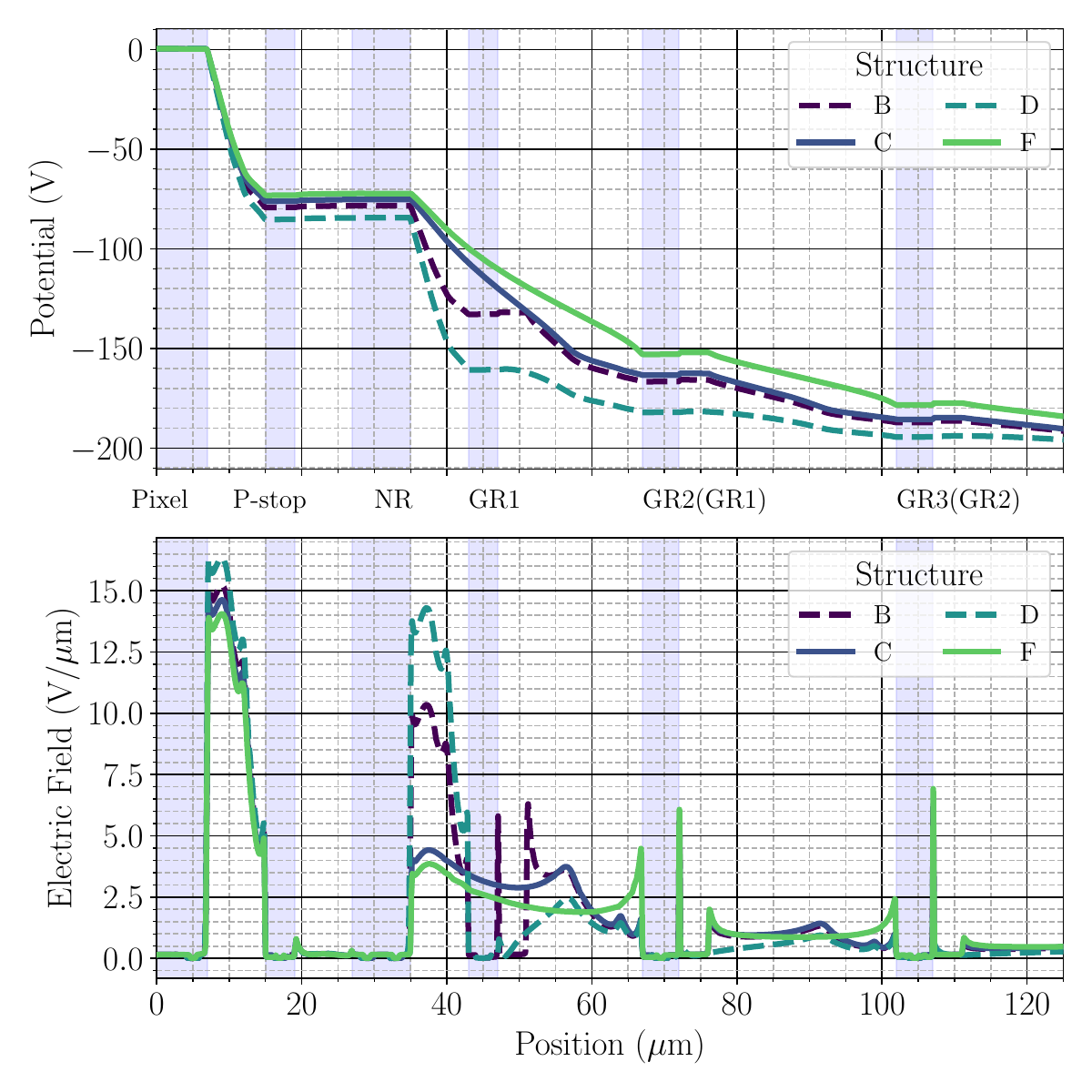}
	\caption{Potential (upper) and electric field strength (lower) distribution of structure B, C, D, and F with a floating n-ring and $V_\mathrm{bias}=-\SI{200}{\volt}$. The region from the pixel to the inner floating guard rings is displayed to focus on the region of the largest potential-drop and maximum electric field. Due to the difference in the guard ring designs, the position of the rings of structures B and D are labelled without parentheses, whereas their labels for structure C and F are listed in the parentheses (See also Figure \ref{GRDetail}). The maximum electric field at the pixel implant is not significantly influenced by the guard ring design.}
	\label{simPotEfStrBCDFfN}
\end{figure}

\begin{figure}[H]
	\centering
	\includegraphics[width=0.48\textwidth]{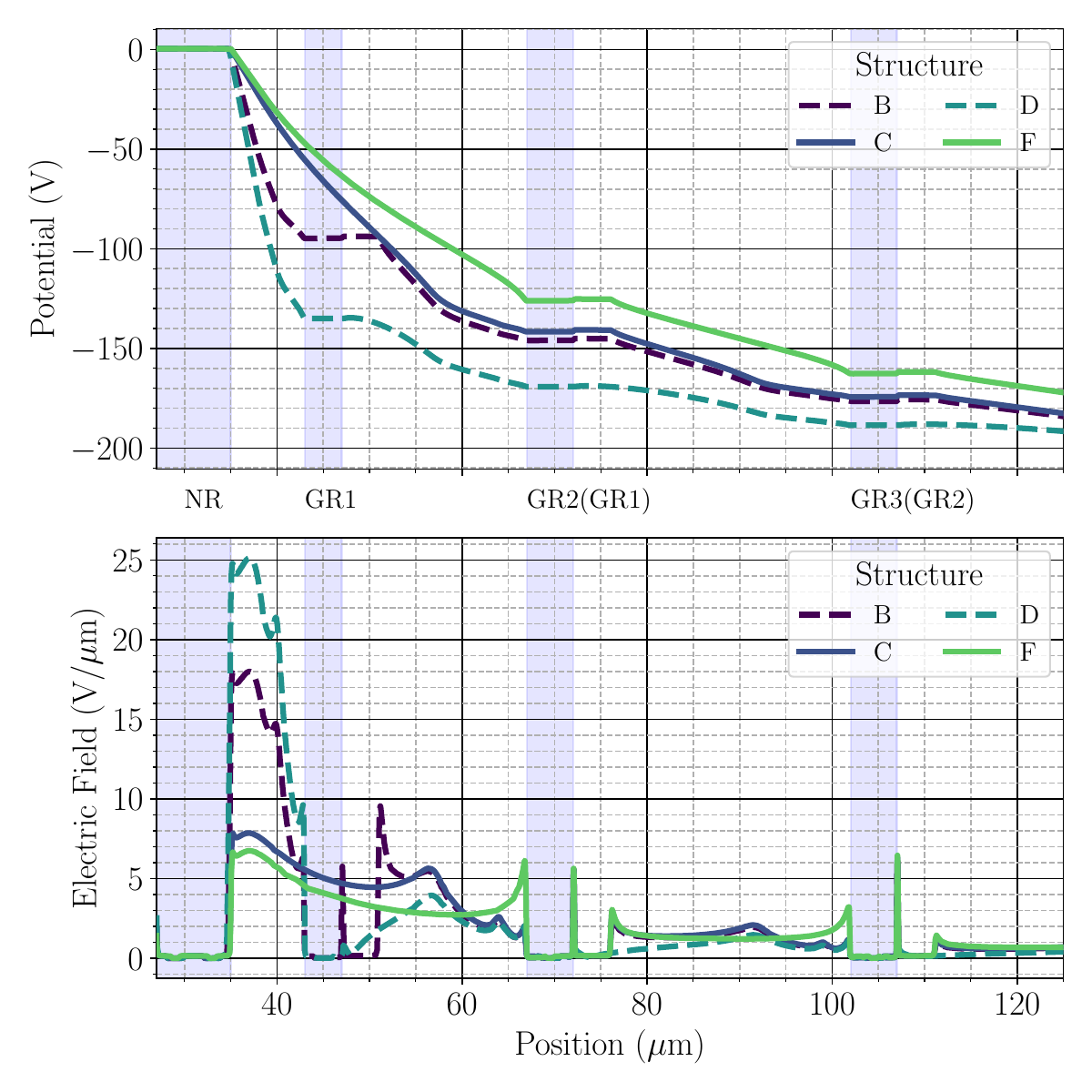}
	\caption{Potential (upper) and electric field strength (lower) distribution of structure B, C, D, and F with a grounded n-ring and $V_\mathrm{bias}=-\SI{200}{\volt}$. Since the maximum potential drop and electric field locates at the n-ring (NR), the region from the NR to the inner floating guard rings is displayed. The labelling of guard ring positions is presented in the same manner as that in Figure \ref{simPotEfStrBCDFfN}. An overall elevation of the floating potential (reduction of the maximum electric field at the NR) is found when the guard rings are equipped with the ``n+p'' implant (``B'' vs. ``D'') and the overhang structure is omitted (``F'' vs. ``C''). Increasing the spacing between the NR and GR1 reduces the slope of the potential distribution at the NR, and reduces the maximum electric field (``C'' vs. ``B'').}
	\label{simPotEfStrBCDFgN}
\end{figure}

For a given bias voltage (e.g.~$V_{\mathrm{bias}}=-\SI{200}{\volt}$), the potentials at the floating wells were simulated to be at different negative potentials. The floating potential is lower when the distance from the guard ring implant to the pixel/n-ring is larger. This behaviour resulted in a gradual decrease of the potential from the pixel matrix to the sensor edge. In both guard ring structures, the maximum potential-drop and subsequently the maximum electric field occurred between pixel and p-stop for a floating n-ring, or n-ring and GR1 for a grounded n-ring. The floating n-ring and the p-stop acted as extra guard rings with a similar potential starting from the pixel. The addition of DNW to the n-ring (as in structure D) elevated the potential at the p-stop, hence reduced the maximum potential-drop and, subsequently, the maximum electric field. This can explain the higher breakdown voltage of structure D in comparison with structure A with a floating n-ring. By grounding the n-ring in structure D, the potential at the p-stop is higher than that in the floating n-ring scenario. Moreover, the maximum electric field is shifted to the edge of the n-ring, and increased by approximately 50\%. As a result, the breakdown voltage of structure D is reduced after grounding the n-ring. For structure A, although the maximum electric field is relocated in the same manner by grounding the n-ring, the field strength is similar as in the case where it is floating. Therefore, the breakdown voltage was similar for both scenarios.\\
\indent
Similar behaviours of the potential and electric field distributions were observed (as depicted in Figure \ref{simPotEfStrBCDFfN}) in structures B, C, and F with a floating n-ring, where the most prominent potential drop and the maximum electric field appears between the pixel and the p-stop. 
With additional NW attached to the guard rings (structure B), the potentials at the n-ring and guard rings were elevated in comparison with the structure without NW (structure D). The increase in the spacing between the n-ring and GR1 (structures C and F) further smoothed the potential distribution in this region, and resulted in a smaller electric field. However, such changes in the guard ring designs only caused small reductions of the maximum electric field. These effects got more pronounced by grounding the n-ring, as can be observed in Figure \ref{simPotEfStrBCDFgN}. 
The elevation of the potentials at the guard rings and the smoothening of the potential drop visibly reduced the maximum electric field which was relocated to the region between the n-ring and GR1. Consequently, the structure C revealed a significantly higher breakdown voltage than the structures B and D.\\
\newpage
The effect of the overhang structure can be more clearly identified through comparing structure C and F with a grounded n-ring, as depicted in Figure \ref{simPotEfStrBCDFgN}. In this case the overhang in structure C causes an overall suppression of the potential at the floating guard rings. Therefore, the maximum electric field at the n-ring of structure C was higher than that in structure F. This ensures a higher breakdown voltage of structure F.\\
\begin{figure}[t]
	\centering
	\includegraphics[width=0.48\textwidth]{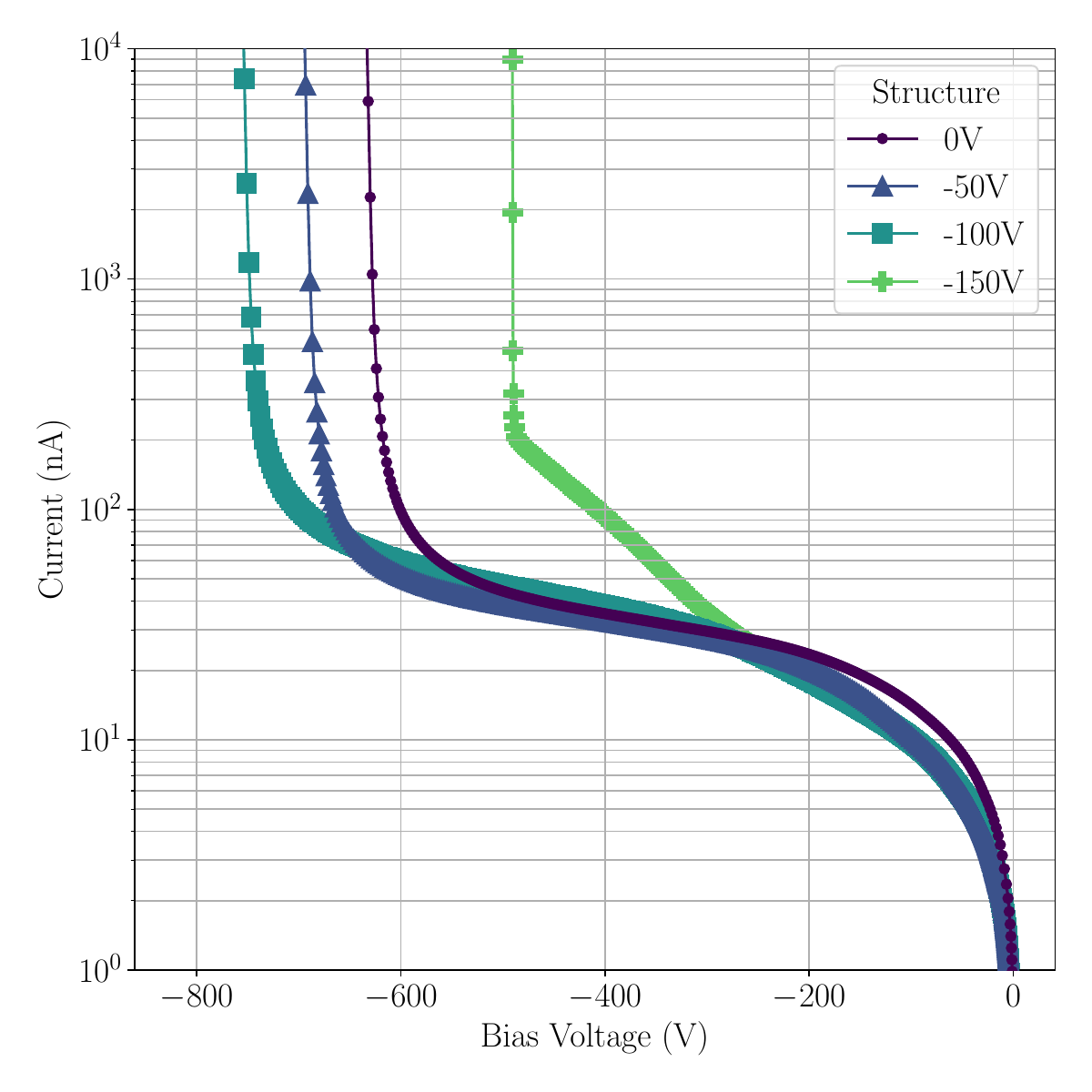}
	\caption{Simulated IV curve for structure E with floating n-ring. The proportionality between $\Delta V_\mathrm{poly}$ and $\Delta V_\mathrm{BD}$, and the early breakdown for $V_\mathrm{poly}=-\SI{150}{\volt}$ are reproduced.}
	\label{simBDIVstrEfN}
\end{figure}
\begin{figure}[t]
	\centering
	\includegraphics[width=0.48\textwidth]{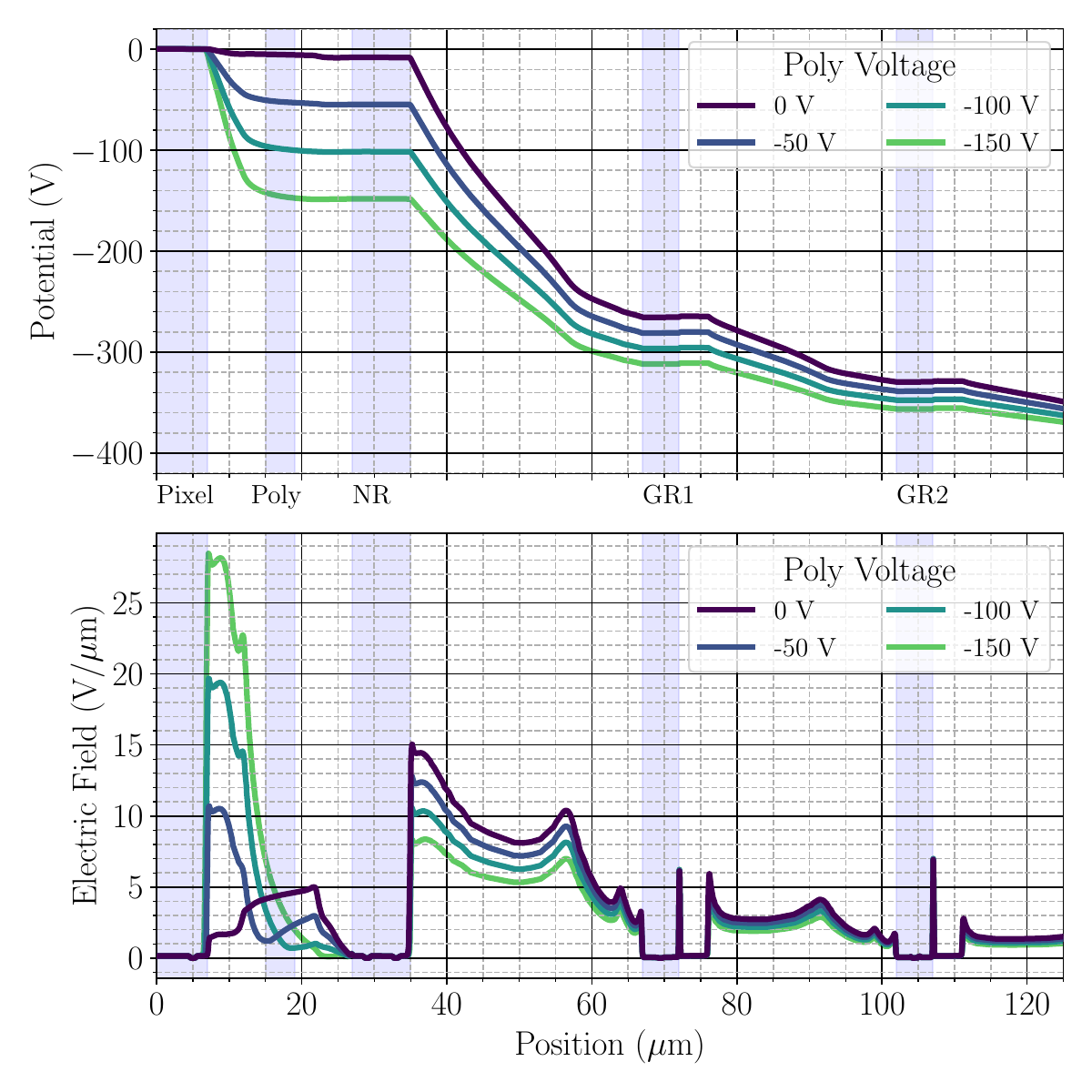}
	\caption{Potential (upper) and electric field strength (lower) distribution of structure E with floating n-ring and $V_\mathrm{bias}=-\SI{400}{\volt}$ for various $V_\mathrm{poly}$. The potential beneath the field-plate is pinned to $V_\mathrm{poly}$ for sufficiently high bias voltages. The increasing $\vert V_\mathrm{poly}\vert$ results in a reduction of electric field at the NR, so that higher bias voltage is required to trigger the breakdown there. For $V_\mathrm{poly}=-\SI{150}{\volt}$, the electric field at the pixel becomes sufficiently high for the onset of breakdown.}
	\label{simPotEfStrEfN}
\end{figure}
\indent
he simulated IV-curves and distributions of structure E with floating n-ring are presented in Figure \ref{simBDIVstrEfN}. According to the extracted breakdown voltages listed in Table \ref{simBDIV200}, the simulation qualitatively reproduced the measurements in 
\ref{200BDfNR}. The early breakdown for $V_{\mathrm{poly}} = -150~\mathrm{V}$ was predicted by the simulation, as well. As depicted in Figure \ref{simPotEfStrEfN}, the potential at the region underneath the field-plate was affected by the set $V_{\mathrm{poly}}$. The field-plate, the underlying silicon oxide, and the silicon substrate forms an MOS structure, where the field-plate acts as the gate (see Figure \ref{InterPixel}(c)). The negative voltage applied to the field-plate determines the potential at the surface of the silicon substrate. During the ramp of bias voltage, the potential in this region was lowered and finally stabilised at approximately $V_{\mathrm{poly}}$. For $V_{\mathrm{poly}}=\SI{0}{\volt}$, the maximum electric field was located at the n-ring, which indicates the location of the junction breakdown. As $\vert V_{\mathrm{poly}}\vert$ was increased, the electric field was reduced at the n-ring and increased at the pixel implant. Adding the bias voltage increased the potential difference and the electric field mainly between the n-ring and GR1 until the onset of breakdown occurs. This gave the result, that the breakdown voltage was lower for $V_{\mathrm{poly}}=\SI{0}{\volt}$ than $V_{\mathrm{poly}}=\SI{-100}{\volt}$, due to the higher electric field at the n-ring. Nevertheless, when the potential beneath the field-plate reached approximately $V_{\mathrm{poly}}=\SI{-150}{\volt}$, the electric field was already large enough to trigger the junction breakdown at the pixel implant. Thus the onset of the breakdown occurred at the pixel for a lower bias voltage, in comparison with other $V_{\mathrm{poly}}$ settings. By grounding the n-ring, the breakdown voltages revealed a less significant response to the change in the $V_{\mathrm{poly}}$ (Table \ref{simBDIV200}). In this case, the maximum electric field was located at the n-ring, which was predominately determined by the bias voltage and less influenced by the potential at the field-plate.\\
\indent
It should be noticed that the results in this work were obtained and discussed based on unirradiated sensors, so that they cannot be directly extrapolated to the cases of radiation-damaged samples. Especially for the effects of the overhang structure and field-plate, where MOS structures are involved, the results can be substantially different, due to the radiation-induced changes on the electrical properties at the ${\mathrm{SiO}_{2}}$-$\mathrm{Si}$ interface. Therefore, the behaviour of the breakdown performance of guard ring structures after considering radiation damages requires a dedicated study.

\section{Conclusion}
Measurements of the current--voltage behaviour were performed on unirradiated passive CMOS test structures fabricated on high resistivity p-type substrate for characterising their breakdown performances. By using 2D TCAD simulations of the guard ring region, the relation between multiple design features and the breakdown voltage were qualitatively reproduced, and discussed based on the simulated potential and electric field distribution. These comparative studies showed that geometrical layouts, implant types, and overhang structures of the floating guard rings determine the strength of the maximum electric field in the guard ring region, and hence the breakdown voltage. From the perspective of the implant geometry and type, the breakdown voltage of an unirradiated sensor can be improved by:
\begin{itemize}
 \item \textit{An increase in the spacing between the n-ring and the first guard ring:}

 A large spacing reduces the steepness of the maximum potential-drop in the guard ring region.
 \item \textit{The use of adjacent n-type and p-type implants as guard rings:}

 Attaching an n-well to the p-type guard ring implant elevates the potential of the floating guard ring, thus smoothing the potential distribution and reducing the maximum electric field. This effect can be further enhanced by the use of deep n-wells.
\end{itemize}
The use of an overhang structure and the field-plate exhibits the modification to the potential distribution via MOS structures. It is found that implementing the overhang structure considered in this work can cause an increase in the maximum electric field, and hence, deliver a worse breakdown performance of unirradiated sensors. Assigning different voltages at the poly silicon field-plate can be used to modify the breakdown voltage, since it can control the potential in the corresponding region and alter the maximum electric field in the sensor. Moreover, future studies are foreseen to explore the implementation of the deep n-wells and the field-plates in the guard ring structure to achieve further improvements of the breakdown performances.

\section{Acknowledgements}
This work has received funding from the German Federal Ministry of Education and Research BMBF (grant 05H15PDCA9) and the European Union's Horizon 2020 Research and Innovation programme under grant agreements No. 675587 (ITN-STREAM), 654168 (AIDA-2020) and 101004761 (AIDA-Innova).

\bibliographystyle{elsarticle-num}
\bibliography{MAIN}

\begin{thebibliography}{10}
\expandafter\ifx\csname url\endcsname\relax
  \def\url#1{\texttt{#1}}\fi
\expandafter\ifx\csname urlprefix\endcsname\relax\def\urlprefix{URL }\fi
\expandafter\ifx\csname href\endcsname\relax
  \def\href#1#2{#2} \def\path#1{#1}\fi

\bibitem{Baliga:2019vi}
B.~J. Baliga, {Fundamentals of Power Semiconductor Devices}, {Springer}, 2019.

\bibitem{BISCHOFF199327}
A.~Bischoff, et~al., Breakdown protection and long-term stabilisation for
  si-detectors, Nucl. Instrum. Methods Phys. Res. A 326~(1) (1993) 27--37.
\newblock \href {http://dx.doi.org/10.1016/0168-9002(93)90329-G}
  {\path{doi:10.1016/0168-9002(93)90329-G}}.

\bibitem{AVSET1996397}
B.~S. Avset, L.~Evensen, {The effect of metal field plates on multiguard
  structures with floating p+ guard rings}, Nucl. Instrum. Methods Phys. Res. A
  377~(2) (1996) 397--403.
\newblock \href {http://dx.doi.org/10.1016/0168-9002(96)00194-5}
  {\path{doi:10.1016/0168-9002(96)00194-5}}.

\bibitem{672633}
N.~Bacchetta, et~al., Study of breakdown effects in silicon multi-guard
  structures, Vol. vol. 1 of 1997 IEEE Nuclear Science Symposium Conference
  Record, 1997, pp. 498--502.
\newblock \href {http://dx.doi.org/10.1109/NSSMIC.1997.672633}
  {\path{doi:10.1109/NSSMIC.1997.672633}}.

\bibitem{rossi2006pixel}
L.~Rossi, P.~Fischer, T.~Rohe, N.~Wermes, {Pixel Detectors: From Fundamentals
  to Applications}, Springer Science \& Business Media, 2006.

\bibitem{2009}
M.~Benoit, A.~Lounis, N.~Dinu, {Simulation of guard ring influence on the
  performance of {ATLAS} pixel detectors for inner layer replacement}, J.
  Instrum. 4~(03) (2009) P03025.
\newblock \href {http://dx.doi.org/10.1088/1748-0221/4/03/p03025}
  {\path{doi:10.1088/1748-0221/4/03/p03025}}.

\bibitem{BORTOLETTO1999178}
D.~Bortoletto, et~al., Radiation damage studies of multi-guard ring p-type bulk
  diodes, Nucl. Instrum. Methods Phys. Res. A 435~(1) (1999) 178--186.
\newblock \href {http://dx.doi.org/10.1016/S0168-9002(99)00438-6}
  {\path{doi:10.1016/S0168-9002(99)00438-6}}.

\bibitem{EGOROV1999197}
N.~Egorov, et~al., Operation of guard rings on the ohmic side of n+--p--p+
  diodes, Nucl. Instrum. Methods Phys. Res. A 426~(1) (1999) 197--205.
\newblock \href {http://dx.doi.org/10.1016/S0168-9002(98)01492-2}
  {\path{doi:10.1016/S0168-9002(98)01492-2}}.

\bibitem{5603386}
O.~{Koybasi}, G.~{Bolla}, D.~{Bortoletto}, {Guard Ring Simulations for n-on-p
  Silicon Particle Detectors}, IEEE Transactions on Nuclear Science 57~(5)
  (2010) 2978--2986.
\newblock \href {http://dx.doi.org/10.1109/TNS.2010.2063439}
  {\path{doi:10.1109/TNS.2010.2063439}}.

\bibitem{GALLRAPP201229}
C.~Gallrapp, et~al., Performance of novel silicon n-in-p planar pixel sensors,
  Nucl. Instrum. Methods Phys. Res. A 679 (2012) 29--35.
\newblock \href {http://dx.doi.org/10.1016/j.nima.2012.03.029}
  {\path{doi:10.1016/j.nima.2012.03.029}}.

\bibitem{Pohl_2017}
D.-L. Pohl, et~al., {Radiation hard pixel sensors using high-resistive wafers
  in a 150 nm {CMOS} processing line}, J. Instrum. 12~(06) (2017) P06020.
\newblock \href {http://dx.doi.org/10.1088/1748-0221/12/06/p06020}
  {\path{doi:10.1088/1748-0221/12/06/p06020}}.

\bibitem{DIETER2021165771}
Y.~Dieter, et~al., {Radiation tolerant, thin, passive {CMOS} sensors read out
  with the {RD53A} chip}, Nucl. Instrum. Methods Phys. Res. A 1015.
\newblock \href {http://dx.doi.org/10.1016/j.nima.2021.165771}
  {\path{doi:10.1016/j.nima.2021.165771}}.

\bibitem{HIRONO201987}
T.~Hirono, et~al., {Depleted fully monolithic active {CMOS} pixel sensors
  ({DMAPS}) in high resistivity 150 nm technology for {LHC}}, Nucl. Instrum.
  Methods Phys. Res. A 924 (2019) 87--91.
\newblock \href {http://dx.doi.org/10.1016/j.nima.2018.10.059}
  {\path{doi:10.1016/j.nima.2018.10.059}}.

\bibitem{Barbero_2020}
M.~Barbero, et~al., {Radiation hard {{DMAPS}} pixel sensors in 150 nm {CMOS}
  technology for operation at {LHC}}, J. Instrum. 15~(05) (2020) P05013.
\newblock \href {http://dx.doi.org/10.1088/1748-0221/15/05/p05013}
  {\path{doi:10.1088/1748-0221/15/05/p05013}}.

\bibitem{Garcia_Sciveres_2018}
M.~Garcia-Sciveres, N.~Wermes, {A review of advances in pixel detectors for
  experiments with high rate and radiation}, Reports on Progress in Physics
  81~(6).
\newblock \href {http://dx.doi.org/10.1088/1361-6633/aab064}
  {\path{doi:10.1088/1361-6633/aab064}}.

\bibitem{lfoundry}
\href{http://www.lfoundry.com/en}{Lfoundry}.
\newline\urlprefix\url{http://www.lfoundry.com/en}

\bibitem{tcad}
\href{https://www.synopsys.com/silicon/tcad.html}{{Synopsys {TCAD}}}.
\newline\urlprefix\url{https://www.synopsys.com/silicon/tcad.html}

\bibitem{CHEN20051311}
S.-H. Chen, M.-D. Ker, {Investigation on seal-ring rules for IC product
  reliability in 0.25-$\mathrm{\mu m}$ {CMOS} technology}, Microelectron.
  Reliabil. 45~(9) (2005) 1311--1316.
\newblock \href {http://dx.doi.org/10.1016/j.microrel.2005.07.012}
  {\path{doi:10.1016/j.microrel.2005.07.012}}.

\bibitem{Hemperek:2018ut}
T.~Hemperek,
  \href{https://nbn-resolving.org/urn:nbn:de:hbz:5n-50354}{{Exploration of
  Advanced {CMOS} Technologies for New Pixel Detector Concepts in High Energy
  Physics}}, Ph.D. thesis, {Rheinischen Friedrich-Wilhelms-Universit\"at Bonn}
  (2018).
\newline\urlprefix\url{https://nbn-resolving.org/urn:nbn:de:hbz:5n-50354}

\bibitem{sze2007physics}
S.~M. Sze, K.~K. Ng, {Physics of Semiconductor Devices}, {John Wiley \& Sons,
  Inc}, 2007.

\bibitem{Baselga_2018}
M.~Baselga, et~al., {Front-side biasing of n-in-p silicon strip detectors},
  Journal of Instrumentation 13~(11) (2018) P11007.
\newblock \href {http://dx.doi.org/10.1088/1748-0221/13/11/p11007}
  {\path{doi:10.1088/1748-0221/13/11/p11007}}.

\bibitem{1474639}
A.~Grove, O.~Leistiko, W.~Hooper, {Effect of surface fields on the breakdown
  voltage of planar silicon p-n junctions}, IEEE Transactions on Electron
  Devices 14~(3) (1967) 157--162.
\newblock \href {http://dx.doi.org/10.1109/T-ED.1967.15916}
  {\path{doi:10.1109/T-ED.1967.15916}}.

\bibitem{sentaurusProcess}
Synopsys, \href{www.synopsys.com}{Sentaurus{\texttrademark} process user
  guide}, www.synopsys.com (September 2020).
\newline\urlprefix\url{www.synopsys.com}

\bibitem{sentaurusDevice}
Synopsys, \href{www.synopsys.com}{Sentaurus{\texttrademark} device user guid},
  www.synopsys.com (September 2020).
\newline\urlprefix\url{www.synopsys.com}

\bibitem{1483108}
G.~Masetti, M.~Severi, S.~Solmi, {Modeling of carrier mobility against carrier
  concentration in arsenic-, phosphorus-, and boron-doped silicon}, IEEE
  Transactions on Electron Devices 30~(7) (1983) 764--769.
\newblock \href {http://dx.doi.org/10.1109/T-ED.1983.21207}
  {\path{doi:10.1109/T-ED.1983.21207}}.

\bibitem{1478102}
C.~Canali, et~al., {Electron and hole drift velocity measurements in silicon
  and their empirical relation to electric field and temperature}, IEEE
  Transactions on Electron Devices 22~(11) (1975) 1045--1047.
\newblock \href {http://dx.doi.org/10.1109/T-ED.1975.18267}
  {\path{doi:10.1109/T-ED.1975.18267}}.

\bibitem{PhysRev.77.388}
E.~Conwell, V.~F. Weisskopf, {Theory of Impurity Scattering in Semiconductors},
  Phys. Rev. 77 (1950) 388--390.
\newblock \href {http://dx.doi.org/10.1103/PhysRev.77.388}
  {\path{doi:10.1103/PhysRev.77.388}}.

\bibitem{Van_Overstraeten_1970}
R.~Van~Overstraeten, H.~De~Man, {Measurement of the ionization rates in
  diffused silicon p-n junctions}, Solid-State Electronics 13~(5) (1970)
  583--608.
\newblock \href {http://dx.doi.org/10.1016/0038-1101(70)90139-5}
  {\path{doi:10.1016/0038-1101(70)90139-5}}.

\bibitem{Shockley:1952aa}
W.~Shockley, {Statistics of the Recombinations of Holes and Electrons},
  Physical Review 87~(5) (1952) 835--842.
\newblock \href {http://dx.doi.org/10.1103/PhysRev.87.835}
  {\path{doi:10.1103/PhysRev.87.835}}.

\bibitem{PhysRev.87.387}
R.~N. Hall, {Electron-Hole Recombination in Germanium}, Physical Review 87
  (1952) 387.
\newblock \href {http://dx.doi.org/10.1103/PhysRev.87.387}
  {\path{doi:10.1103/PhysRev.87.387}}.

\bibitem{121690}
G.~A.~M. {Hurkx}, D.~B.~M. {Klaassen}, M.~P.~G. {Knuvers}, {A new recombination
  model for device simulation including tunneling}, IEEE Transactions on
  Electron Devices 39~(2) (1992) 331--338.
\newblock \href {http://dx.doi.org/10.1109/16.121690}
  {\path{doi:10.1109/16.121690}}.

\bibitem{Morozzi_2021}
A.~Morozzi, et~al., {{TCAD} Modeling of Surface Radiation Damage Effects: {A}
  State-Of-The-Art Review}, Front. Phys. 9.
\newblock \href {http://dx.doi.org/10.3389/fphy.2021.617322}
  {\path{doi:10.3389/fphy.2021.617322}}.

\bibitem{Hirono:2019aa}
T.~Hirono, \href{http://hdl.handle.net/20.500.11811/7933}{{Development of
  Depleted Monolithic Active Pixel Sensors for High Rate and High Radiation
  Experiments at HL-LHC}}, Ph.D. thesis, {Rheinischen
  Friedrich-Wilhelms-Universit\"at Bonn} (2019).
\newline\urlprefix\url{http://hdl.handle.net/20.500.11811/7933}

\bibitem{Wang_2020}
T.~Wang, et~al., {Depleted Monolithic Active Pixel Sensors in the LFoundry 150
  nm and {TowerJazz} 180 nm {CMOS} Technologies}, Sissa Medialab, 2020.
\newblock \href {http://dx.doi.org/10.22323/1.373.0026}
  {\path{doi:10.22323/1.373.0026}}.

\bibitem{Wang_2018}
T.~Wang, et~al., {Depleted fully monolithic {CMOS} pixel detectors using a
  column based readout architecture for the {ATLAS} Inner Tracker upgrade}, J.
  Instrum. 13~(03) (2018) C03039.
\newblock \href {http://dx.doi.org/10.1088/1748-0221/13/03/c03039}
  {\path{doi:10.1088/1748-0221/13/03/c03039}}.

\bibitem{IGUAZ2019652}
F.~Iguaz, et~al., {Characterization of a depleted monolithic pixel sensors in
  150 nm {CMOS} technology for the {ATLAS} Inner Tracker upgrade}, Nucl.
  Instrum. Methods Phys. Res. A 936 (2019) 652--653.
\newblock \href {http://dx.doi.org/10.1016/j.nima.2018.11.009}
  {\path{doi:10.1016/j.nima.2018.11.009}}.

\bibitem{DINGFELDER2022166747}
J.~Dingfelder, et~al., {Progress in {DMAPS} developments and first tests of the
  {{Monopix2}} chips in 150 nm LFoundry and 180 nm {TowerJazz} technology},
  Nucl. Instrum. Methods Phys. Res. A 1034.
\newblock \href {http://dx.doi.org/10.1016/j.nima.2022.166747}
  {\path{doi:10.1016/j.nima.2022.166747}}.

\bibitem{Caicedo:2023aa}
I.~Caicedo, et~al., {Improvement in the Design and Performance of the Monopix2
  Reticle-Scale {DMAPS}}, in: JPS Conf. Proc., 2023, proceedings of the 31st
  International Workshop on Vertex Detectors (in Press).

\end{thebibliography}
\end{document}